\documentclass[fleqn,10pt]{wlscirep}
\usepackage[utf8]{inputenc}
\usepackage[T1]{fontenc}

\usepackage{graphicx}
\usepackage{dcolumn}
\usepackage{bm}
\usepackage{siunitx}
\usepackage{gensymb}

\usepackage{amsmath} 

\usepackage[caption=false]{subfig}
\usepackage{xcolor}
\usepackage{hyperref}
\usepackage{hyperref}
\hypersetup{
    colorlinks=true,
    linkcolor=blue,
    filecolor=blue,      
    urlcolor=blue,
    citecolor=blue,
}
 
\newcommand{\units}[1]{\ensuremath{\,\mathrm{#1}}}
\DeclareUnicodeCharacter{2212}{-}

\title{Pt and CoB trilayer Josephson $\pi$ junctions with \\perpendicular magnetic anisotropy}

\author[1]{N.~Satchell}
\author[1]{T.~Mitchell}
\author[1]{P. M.~Shepley}
\author[1]{E.~Darwin}
\author[1]{B. J.~Hickey}
\author[1,*]{G.~Burnell}
\affil[1]{School of Physics and Astronomy, University of Leeds, Leeds, LS2 9JT, United Kingdom}
\affil[*]{g.burnell@leeds.ac.uk}

\begin{abstract}

  We report on the electrical transport properties of Nb based Josephson junctions with Pt/Co$_{68}$B$_{32}$/Pt ferromagnetic barriers. The barriers exhibit perpendicular magnetic anisotropy, which has the main advantage for potential applications over magnetisation in-plane systems of not affecting the Fraunhofer response of the junction. In addition, we report that there is no magnetic dead layer at the Pt/Co$_{68}$B$_{32}$ interfaces, allowing us to study barriers with ultra-thin Co$_{68}$B$_{32}$. In the junctions, we observe that the magnitude of the critical current oscillates with increasing thickness of the Co$_{68}$B$_{32}$ strong ferromagnetic alloy layer. The oscillations are attributed to the ground state phase difference across the junctions being modified from zero to $\pi$. The multiple oscillations in the thickness range $0.2~\leqslant~d_\text{CoB}~\leqslant~1.4$~nm suggests that we have access to the first zero-$\pi$ and $\pi$-zero phase transitions. Our results fuel the development of low-temperature memory devices based on ferromagnetic Josephson junctions.

\end{abstract}
\begin{document}

\flushbottom
\maketitle
%
%
\thispagestyle{empty}

\section*{Introduction}

Proximity effects between superconducting (\textit{S}) and ferromagnetic (\textit{F}) materials are a topic of intense research effort due to the new physics at \textit{S--F} interfaces \cite{RevModPhys.77.935, RevModPhys.77.1321, eschrig_spin-polarized_2011, linder_superconducting_2015, 0034-4885-78-10-104501,doi:10.1098/rsta.2015.0150}. In \textit{S--F--S} Josephson junctions, it is well established that the ground-state phase difference across the junction can be tuned from zero to $\pi$, depending on the \textit{F} layer thickness \cite{RevModPhys.77.935}. Experimentally, the zero-$\pi$ transitions correspond to oscillations in the junction's characteristic voltage, $I_cR_N$, with increasing \textit{F} layer thickness \cite{buzdin1982critical}. To date, experimental demonstrations of $\pi$-junctions include: the weak ferromagnetic alloys CuNi \cite{PhysRevLett.86.2427,PhysRevB.68.054531,PhysRevLett.96.197003,PhysRevLett.97.247001,doi:10.1063/1.2356104,PhysRevLett.121.177702,bolginov2018fabrication}, PdNi \cite{PhysRevLett.89.137007,PhysRevB.79.094523} and PdFe \cite{7747519}, the ferromagnetic elements Ni \cite{PhysRevLett.89.187004,PhysRevB.73.174506,PhysRevLett.97.177003,PhysRevB.76.094522,4277679,PhysRevB.79.054501,baek2014hybrid,PhysRevApplied.7.064013,8359359}, Co \cite{PhysRevLett.97.177003,PhysRevB.76.094522,4277679,doi:10.1063/1.3262969} and Fe \cite{PhysRevLett.97.177003,piano20070,PhysRevB.76.094522}, and the strong ferromagnetic alloys NiFe \cite{PhysRevB.71.180501,PhysRevLett.97.177003,PhysRevB.76.094522,4277679,doi:10.1063/1.4862195,doi:10.1063/1.4989392,PhysRevB.97.024517}, Ni$_3$Al \cite{PhysRevB.74.140501}, NiFeMo \cite{Niedzielski_2015} and NiFeCo \cite{doi:10.1063/1.4989392}.

In general, most previous works measure Josephson junctions with \textit{F} layers having in-plane magnetisation. When the magnetisation is in-plane, the \textit{F} layer can contribute significant magnetic flux density in the junction, modifying the response of the junction to an externally applied measurement field and shifting the maximum critical current away from $H=0$. In addition, when applying an in-plane field to measure the junction, the \textit{F} layer may switch in the measurement field. Josephson junctions containing perpendicular magnetic anisotropy (PMA) \textit{F} layers have advantages over in-plane systems as, in principle, the magnetisation and magnetic switching of layers in the junction should not affect the in-plane magnetic flux. The application of an in-plane measurement field will tilt the magnetisation of the PMA layer slightly, however the field required to fully saturate the PMA layers considered in this work is far larger than the field required to characterise the junctions, so the tilting effect can be neglected.

Of the previous \textit{F} layers characterised, only CuNi and PdNi have an intrinsic PMA component of their magnetisation. An alternative to intrinsic PMA is interfacial PMA, which can give a \textit{F} layer an overall PMA so long as the \textit{F} layer is thin enough that the interfacial anisotropy dominates over the bulk anisotropy. Josephson junctions containing interfacial PMA \textit{F} layers have been previously studied theoretically\cite{Margaris_2010,PhysRevB.92.174516,PhysRevB.95.184508} and experimentally in the context of spin-triplet supercurrents, however, no zero-$\pi$ oscillations were expected or observed in the particular geometries studied \cite{PhysRevB.86.224506, PhysRevB.96.224515, Glickeaat9457, PhysRevB.99.174519}.

In this work, we study the amorphous strong ferromagnetic alloy Co$_{68}$B$_{32}$ \cite{PhysRevB.47.2671}. For many spintronic applications, the amorphous Co based alloys are advantageous over crystalline Co due to their lack of crystalline anisotropy and weaker pinning of magnetic domain walls due to the reduced density of grain boundaries \cite{doi:10.1063/1.3280373}. Recently, thin film Co$_{68}$B$_{32}$ has been studied for magnetic memory application and as a host of magnetic skyrmions \cite{schellekens2012electric,finizio2019deterministic,zeissler2020diameter}. When placed adjacent to Pt layers, the Pt/Co$_{68}$B$_{32}$ interfaces exhibit PMA, giving an overall PMA for the thin layers considered in this work. Previously, we used Co$_{68}$B$_{32}$ in PMA pseudospin-valve junctions, where the critical current of the junction could be controlled by the relative orientation of two ferromagnets in the Pt/Co/Pt/Co$_{68}$B$_{32}$/Pt barrier \cite{SatchellPSV}. For application in cryogenic memory, it is important to demonstrate that in addition to modulating the critical current of such devices, it is also possible to switch such devices from the zero to $\pi$ state \cite{bell_controllable_2004,gingrich2016controllable,dayton2017experimental,Madden_2018}. For this, the zero--$\pi$ critical current oscillations of the component ferromagnets in the pseudospin-valve should be well characterised. In this work, we present evidence of such zero--$\pi$ critical current oscillations in Pt/Co$_{68}$B$_{32}$/Pt junctions.

\section*{Results}

\subsection*{Magnetic Characterisation}

\begin{figure}[ht]
\centering
        \includegraphics[width=0.5\columnwidth]{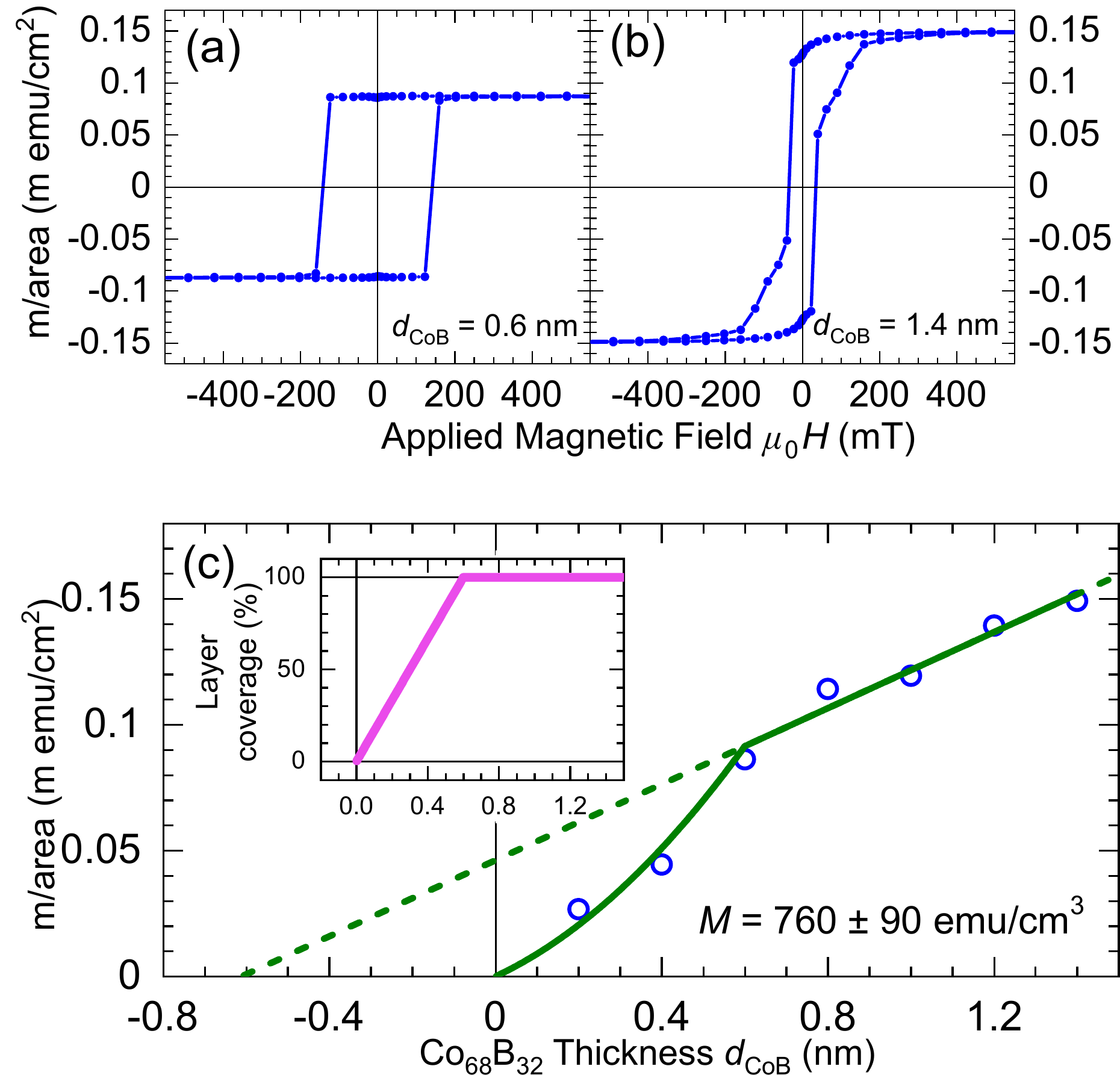}

        \caption{Magnetic characterisation of the sheet film samples \textit{S}-Pt(10)-Co$_{68}$B$_{32}$(\textit{d}$_\text{CoB}$)-Pt(5)-\textit{S}. (a,b) Magnetic hysteresis loops of the magnetic moment (m) per area acquired at a temperature of 10 K with the applied field oriented out-of-plane for (a) \textit{d}$_\text{CoB} = 0.6$ nm and (b) \textit{d}$_\text{CoB} = 1.4$ nm. The diamagnetic contribution from the substrate has been subtracted. (c) Collated saturation moment per area versus nominal thickness of Co$_{68}$B$_{32}$. To model the lower m/area of the thinnest samples in the study, we construct a partial layer coverage model detailed in the text and shown in the inset. The result of fitting the model over the entire data range is shown by the solid line. The extracted magnetisation $M=760 \pm 90$ emu/cm$^3$. The dashed line shows an extrapolation of the linear part of the model to the intercepts. Values of m/area are calculated from the measured total magnetic moments and areas of the samples. The uncertainty in each point is dominated by the area measurements, and is less than 5\%. }
        \label{magnetics}
\end{figure}

Magnetic moment per area versus out-of-plane field data are shown in Figure \ref{magnetics} (a,b) for \textit{S}-Pt(10)-Co$_{68}$B$_{32}$(\textit{d}$_\text{CoB}$)-Pt(5)-\textit{S} sheet film samples at 10~K with a nominal thickness (a) \textit{d}$_\text{CoB} = 0.6$ nm and (b) \textit{d}$_\text{CoB} = 1.4$ nm. For \textit{d}$_\text{CoB} = 0.6$ nm, the square hysteresis loop indicates a strong PMA. As the nominal thickness of the Co$_{68}$B$_{32}$ is increased towards the largest thickness studied in this work, \textit{d}$_\text{CoB} = 1.4$ nm Figure \ref{magnetics} (b), we observe two changing characteristics in the hysteresis loops. Firstly, the coercive field reduces. Secondly, the squareness ratio of the loop reduces - indicating competing anisotropies in the Co$_{68}$B$_{32}$ layer. Upon making the CoB thicker, we would expect that the anisotropy of the layer will change from being predominately PMA to predominately in-plane.

Saturation magnetic moment per area versus nominal thickness of the Co$_{68}$B$_{32}$(\textit{d}$_\text{CoB}$) at 10~K are shown in Figure~\ref{magnetics}~(c). These data are best described in two regimes. For the thicker samples in this study, we observe the expected linear dependence with increasing thickness of ferromagnet. The thinnest two samples in this study deviate from this linear trend, showing a lower moment/area than implied by the trend in the thicker samples. 

In order to apply a fitting model which describes the entire data set, we construct a partial layer coverage toy model of our system. The basis of the model is the assumption that the $F$-layers in the thinnest samples may not be continuous. From zero thickness to some critical thickness, we assume that the layer coverage increases linearly by the profile shown in Figure \ref{magnetics} (c) inset. The physical picture implied by this model is consistent with island nucleation, coalescence followed by layered growth – which is not an atypical growth mode for ambient temperature sputtered thin films. Above the critical thickness, we assume that the layer is now continuous, so the data can be described by the expected linear trend. More information on the model and extracting magnetisation from moment/area data are included in the Supplementary Information along with alternative fitting models to the data presented in Figure \ref{magnetics} (c) (see Supplementary Fig. S1 online).

The result of our toy model is shown by the solid line in Figure \ref{magnetics} (c). The toy model gives a critical thickness for layer growth of $0.6\pm0.1$~nm, which is comparable to a couple of unit cells of nominal thickness, and gives the magnetisation of the Co$_{68}$B$_{32}$ to be $M=760\pm 90$~emu/cm$^3$, consistent with the expected bulk magnetisation of 730~emu/cm$^3$ \cite{312324}. The model also includes a contribution to the total magnetic signal from the polarisation of the adjacent Pt layer, this is commonly observed in such systems  \cite{doi:10.1063/1.344903, PhysRevB.65.020405, PhysRevB.72.054430,  rowan2017interfacial, PhysRevB.100.174418}. In Figure \ref{magnetics} (c), the polarised Pt contribution can be extracted from the y-intercept of the dashed line, which is an extrapolation of the linear part of the model. The magnetic contribution to the total magnetic response of the sample by the polarised Pt is $46 \pm 9~\mu$emu/cm$^2$, or $23 \pm 5~\mu$emu/cm$^2$ per Pt/Co$_{68}$B$_{32}$ interface, consistent with Suzuki \textit{et al.} \cite{PhysRevB.72.054430}

It has been reported elsewhere that significant magnetic dead layers can form in ferromagnetic Josephson junction barriers at the Nb/\textit{F} interfaces, see for example \cite{PhysRevB.76.094522}, however adding buffer layers such as Rh, Cu or Pt can significantly improve the morphology of the \textit{F} layer \cite{doi:10.1063/1.3262969,PhysRevB.80.020506,satchellSOC2018,PhysRevB.99.174519}. The signature of magnetic dead layers is a positive x-intercept when plotting moment/area versus thickness. In our Pt/Co$_{68}$B$_{32}$/Pt barriers, the x-intercept when fitting to Figure~\ref{magnetics}~(c) is not positive (regardless of the model used, see Supplementary Fig. S1 online), suggesting that such dead layers have been minimised by the Pt interfaces. Additionally, when the nominal thickness of the Co$_{68}$B$_{32}$ is equivalent to only one or two monolayers, and so the layer is modelled with partial coverage, the polarised Pt appears to have stabilised the magnetisation of what we expect are islands of Co$_{68}$B$_{32}$, allowing us to measure a magnetic response even for \textit{d}$_\text{CoB} = 0.2$ nm, a significant advantage of our approach.

\subsection*{Electrical Transport}

\begin{figure}
\centering
        \includegraphics[width=0.9\columnwidth]{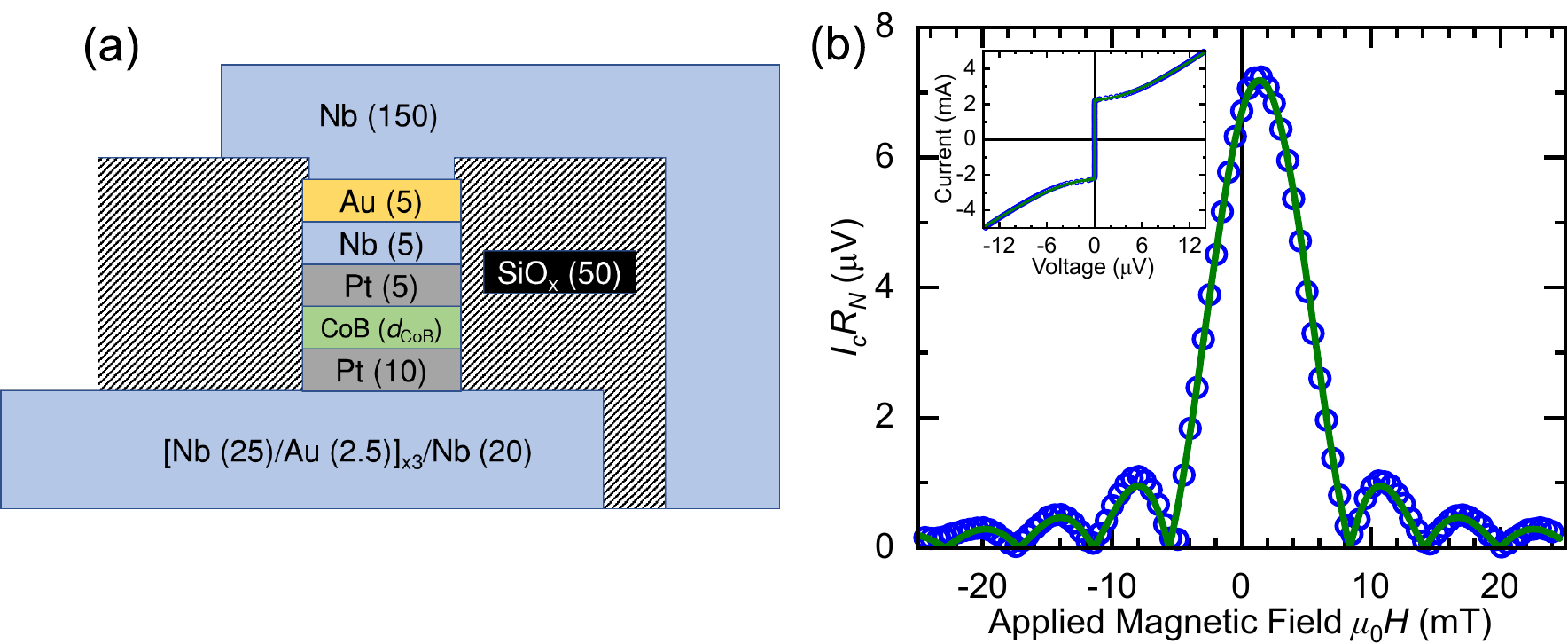}

        \caption{(a) Schematic cross section of the \textit{S}-\textit{F}-\textit{S} Josephson junction device (not to scale). The thickness of each layer is given in nanometers. The Co$_{68}$B$_{32}$ layer thickness, \textit{d}$_\text{CoB}$, is ranged from 0.2 to 1.4~nm. (b) Product of critical Josephson current times normal-state resistance versus applied magnetic field for the device depicted in (a) with \textit{d}$_\text{CoB} = 0.6$~nm at 1.8~K. $I_c$ is determined from the measured $I$--$V$ characteristic at each field value and $R_N$ is the average normal state resistance across all measured fields. The uncertainty in determining $I_cR_N$ is smaller than the data points. The data are fit with Equations \ref{Airy} and \ref{Phi}. Inset: The $I$--$V$ characteristic at zero applied field with fit to Equation \ref{V}. }
        \label{IcB}
\end{figure}

Samples were fabricated using standard lithography techniques into circular current perpendicular-to-plane Josephson junction devices, as depicted schematically in Figure \ref{IcB} (a). We load the devices into our cryostat at room temperature and first cool to 15~K, just above the superconducting transition (9~K), where we apply a 1~T out-of-plane saturating field. Once the saturating field is removed, we rotate the sample so the field is applied in-plane and cool the samples to the base temperature of our cryostat, 1.8~K. We measure the \textit{I--V} characteristic of each junction as a function of in-plane applied magnetic field between $\pm 25$~mT to determine the Fraunhofer pattern. The field necessary to saturate our Pt/Co$_{68}$B$_{32}$/Pt barriers is in excess of 1~T so the maximum deviation of the magnetisation from the perpendicular is less than 1.5$^\circ$.

The \textit{I--V} characteristics of our devices follow the standard square-root form expected for over-damped Josephson junctions \cite{barone1982physics}, 
\begin{equation}
\label{V}
V=R_N\sqrt{I^2 - I^2_{c}},~\text{for}~I \geq I_{c} 
\end{equation}
\noindent where $I_c$ is the critical Josephson current and $R_N$ is the normal state resistance of the junction. For circular Josephson junctions, the $I_c$(\textit{B}) Fraunhofer response can be described by the Airy function \cite{barone1982physics},
\begin{equation}
\label{Airy}
I_c = I_{c0} \left | 2J_1 (\pi \Phi / \Phi_0)/(\pi \Phi / \Phi_0) \right |,
\end{equation}
\noindent where $I_{c0}$ is the maximum critical current, $J_1$ is a Bessel function of the first kind, $\Phi_0=h/2e$ is the flux quantum, and $\Phi$ is the flux through the junction\cite{barone1982physics}, 
\begin{equation}
\label{Phi}
\Phi = \mu_0 (H_\text{app} - H_\text{shift}) w \big[ \lambda^\text{bottom}_\text{L} \tanh{(d^\text{bottom}_S/2\lambda^\text{bottom}_\text{L})} + \lambda^\text{top}_\text{L} \tanh{(d^\text{top}_S/2\lambda^\text{top}_\text{L})} + d \big],
\end{equation}

\noindent where $w$, $\lambda_\text{L}$, $d_S$, and $d$ are the width of the junction, the London penetration depth, the thickness of the superconducting electrode, and the total thickness of all the normal metal layers and \textit{F} layers in the junction, respectively. The bottom electrode is a Nb/Au multilayer ($\lambda^\text{bottom}_\text{L}$ = 190 nm \cite{quarterman2020distortions}) and the top electrode is single layer Nb ($\lambda^\text{top}_\text{L}$ = 150 nm \cite{PhysRevLett.120.247001}). $H_\text{app}$ is the applied field and $H_\text{shift}$ is the amount $I_{c0}$ is shifted from $H$ = 0. $H_\text{shift}$ arises from a combination of an intrinsic contribution due to any in-plane magnetisation of the junction, and extrinsic artifacts from trapped flux in the 3\units{T} superconducting coil used to perform the measurements. Fits to these equations are shown along with the data on a typical device in Figure \ref{IcB} (b). We attribute the small $H_\text{shift}$ in Figure \ref{IcB} (b) to trapped flux in our superconducting coil, which is supported by additional Fraunhofer data for both +1~T and -1~T saturation fields (see Supplementary Fig. S2 online). We determine $I_{c0}$ for many samples of different Co$_{68}$B$_{32}$ thicknesses (\textit{d}$_\text{CoB}$) following the same protocol.

\begin{figure}
\centering
        \includegraphics[width=0.5\columnwidth]{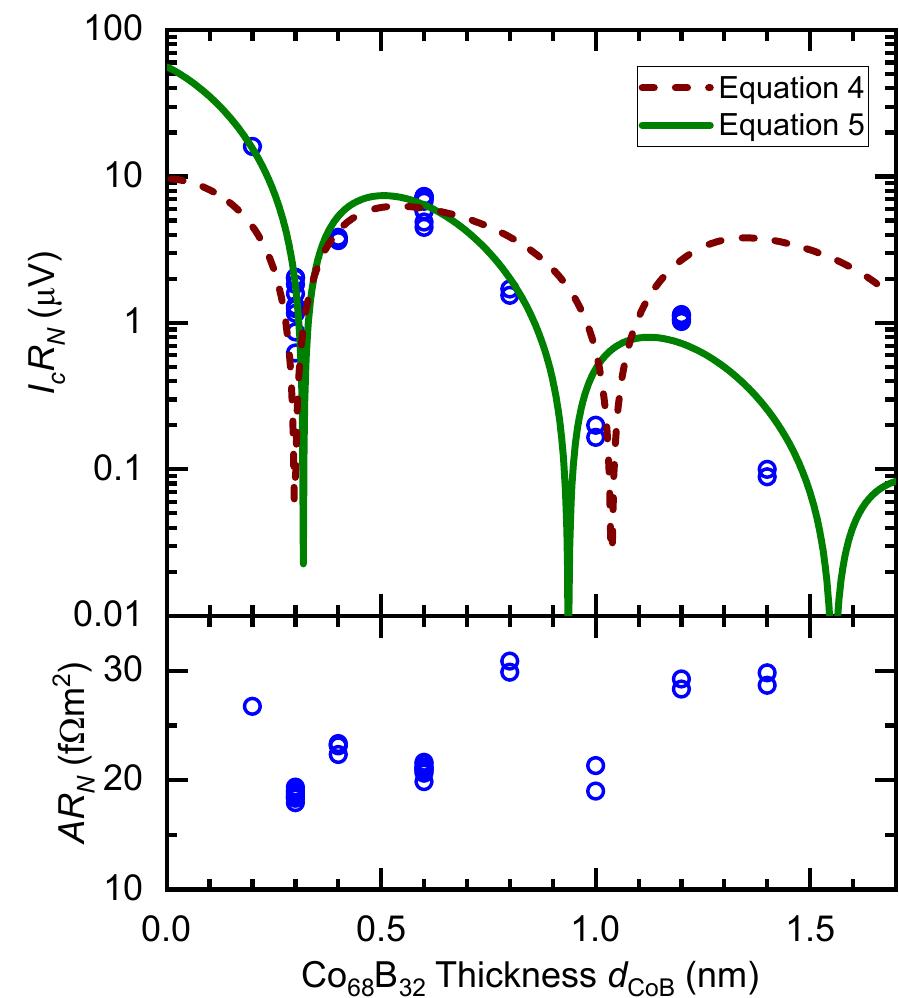}

        \caption{Top: Product of critical Josephson current times normal-state resistance versus nominal Co$_{68}$B$_{32}$ thickness for ferromagnetic Josephson junctions of the form \textit{S}-Pt(10)-Co$_{68}$B$_{32}$(\textit{d}$_\text{CoB}$)-Pt(5)-\textit{S} at 1.8~K. Each data point represents one Josephson junction and the uncertainty in determining $I_cR_N$ is smaller than the data points. The data are fit to Equations \ref{eq:ballistic} and \ref{eq:intermediate}. The best fit parameters for Equation \ref{eq:ballistic} corresponds to $\xi_{F} = 0.28\pm0.01$~nm, and for Equation \ref{eq:intermediate} to $\xi_{F1} = 0.28\pm0.02$~nm and $\xi_{F2} = 0.20\pm0.02$~nm. The first minimum at $0.30\pm0.05$~nm indicates a transition between the zero and $\pi$-phase states. Bottom: Product of the area times normal-state resistance for the same junctions. The scatter in $AR_N$ is most likely sample-to-sample variation in $A$. }
        \label{IcOscillations}
\end{figure}

Figure \ref{IcOscillations} shows the collated $I_cR_N$ and $AR_N$ (area times normal-state resistance) products for the Josephson junctions measured in this study. $I_c$ corresponds to the $I_{c0}$ maximum of the $I_c$(\textit{B}) Fraunhofer response and $R_N$ is the average resistance from measurements at all field values. As the thickness of the Co$_{68}$B$_{32}$ is increased, the $I_cR_N$ shows nonmonotonic behaviour. When plotting $AR_N$, we fix $A$ by the nominal design dimension. The $AR_N$ product for our samples is suggestive that within the same chip the junction-to-junction reproducibility is very good, which is also supported by the small spread of $I_cR_N$ values for junctions on the same chip. It is possible to determine $A$ by fitting $I_c$(\textit{B}) to Equations \ref{Airy} and \ref{Phi}, and we find that across all our junctions the extracted average $\bar{w} = 3.0\pm0.3~\mu$m is consistent with the lithography design. The scatter in $AR_N$ is therefore similar to the scatter in the linear dimensions of the junctions. Variations in $A$ between samples will not affect the reported $I_cR_N$, which is a size independent quantity. Indeed, there is no correlation between a high/low $I_cR_N$ and $AR_N$. 

We report strong reproduciblity of our results, as multiple samples for $d_\text{CoB}= 0.3$ and $0.6$~nm are grown and fabricated in independent cycles and show consistency in $I_cR_N$, Figure \ref{IcOscillations}. Scatter in $I_cR_N$ is most likely driven by sample-to-sample variations in the thickness of the Co$_{68}$B$_{32}$.

\subsection*{Coherence Lengths in \textit{S}/\textit{F}/\textit{S} Josephson Junctions}

The transport properties of \textit{S--F--S} Josephson junctions are well described in three limits, driven by the relative magnitude of three lengthscales; the mean free path ($l_e$), the superconducting coherence length ($\xi_S$) and the effective coherence length inside the ferromagnet ($\xi_F$). In the ballistic limit $l_e > \xi_S > \xi_F$, in the intermediate limit $\xi_S > l_e  > \xi_F$, and in the diffusive limit $\xi_S  > \xi_F > l_e$. 

In the ballistic limit, the decay and oscillations of $I_cR_N$ is given by the numerical maximum with respect to $\varphi$ of the ballistic limit supercurrent $I_\text{S}(\varphi)$ \cite{buzdin1982critical},
\begin{equation}\label{eq:ballistic}
I_\text{S}(\varphi)R_N = \frac{\pi \Delta \alpha^2}{2 e} \int_{\alpha}^{\infty}\frac{dy}{y^3}\Bigg( \sin{}\frac{\varphi-y}{2}\tanh{}\frac{\Delta\cos{}\frac{\varphi-y}{2}}{2k_\text{B}T} + \sin{}\frac{\varphi+y}{2}\tanh{}\frac{\Delta\cos{}\frac{\varphi+y}{2}}{2k_\text{B}T}\Bigg),
\end{equation}
\noindent where $\varphi$ is the phase difference across the junction, $\Delta$ is the energy gap, $T$ is the temperature, and $\alpha \equiv d / \xi_F$. In the ballistic limit, $\xi_F =\hbar v_F /2E_\text{Ex}$, where $v_F$ is the Fermi velocity and $E_\text{Ex}$ is the exchange energy. Ballistic limit transport has been reported in the ferromagnetic elements when sufficiently thin, for example in Ni barriers studied by Robinson \textit{et al.} \cite{PhysRevLett.97.177003} and Baek \textit{et al.} \cite{PhysRevApplied.7.064013}.

In the intermediate limit, the decay and oscillations of $I_cR_N$ is given by \cite{PhysRevB.64.134506},
\begin{equation}\label{eq:intermediate}
I_cR_N = V_0 \exp \bigg( \frac{-d_F}{\xi_{F1}} \bigg) \bigg|\sin\bigg( \frac{d_F - d_{\text{zero--}\pi}}{\xi_{F2}} \bigg)\bigg|,
\end{equation}
\noindent where $d_{\text{zero--}\pi}$ is the thickness of the first zero--$\pi$ transition, $\xi_{F1} = l_e$ and $\xi_{F2}=\xi_{F}$ are the lengthscales governing the decay and oscillation of $I_cR_N$, respectively. In the intermediate limit, one finds $\xi_{F1} > \xi_{F2}$. Most ferromagnetic alloys studied in ferromagnetic Josephson junctions are solid-solutions with short mean free paths and are found to be best described in the intermediate limit\cite{7747519,doi:10.1063/1.4989392}, for example PdNi barriers studied by Khaire \textit{et al.} \cite{PhysRevB.79.094523}.

It is well established that Rashba effects arising from spin-orbit coupling can be found at the interface between metallic ferromagnets, such as Co, and heavy metals, such as Pt\cite{PhysRevB.71.201403,miron2010current,miron2011perpendicular} and so in the diffusive limit with spin-flip or spin-orbit scattering, the transport may be described by Equation \ref{eq:intermediate} \cite{PhysRevB.73.064505}. However in the diffusive limit one will find $\xi_{F2} > \xi_{F1}$ \cite{PhysRevB.73.064505, PhysRevB.84.144513}. In experimental literature, this situation is somewhat rarer, for example CuNi barriers studied by Oboznov \textit{et al.} \cite{PhysRevLett.96.197003} and NiFeMo barriers studied by Niedzielski \textit{et al.} \cite{Niedzielski_2015}.

Fitting to our results taking $\xi_{F}$ and $\xi_{F1,2}$ as fitting parameters (shown in Figure \ref{IcOscillations}), we find that Equation \ref{eq:ballistic} does not reproduce our Co$_{68}$B$_{32}$ data as well as Equation \ref{eq:intermediate}, particularly for larger $d_\text{CoB}$. The fits for Equation \ref{eq:intermediate} correspond to the limit $\xi_{F1} > \xi_{F2}$, placing our junctions in the intermediate limit. The best fit parameters are given in Table \ref{table1}, which also includes results from other ferromagnetic alloys best described by Equation \ref{eq:intermediate}. For further analysis on our data see Supplementary Fig. S3 and S4 online.

\section*{Discussion}

Systems, such as ours, with a source of $s$-wave superconductivity, large spin-orbit coupling, and ferromagnetism are predicted to display transport properties consistent with spin-triplet supercurrents \cite{PhysRevB.89.134517}, reported to be observed in ferromagnetic resonance measurements\cite{jeon2018enhanced}. In this work, however, within the resolution of our measurements, the data are well described by singlet transport physics alone and we do not need to invoke a significant spin-triplet supercurrent to explain our results -- for further details see the Supplementary Information online. The lack of evidence for spin-triplet supercurrents in these Josephson junctions is consistent with previous works \cite{satchellSOC2018,PhysRevB.99.174519}.


In comparison to other ferromagnetic alloys, such as those in Table \ref{table1}, our Co$_{68}$B$_{32}$ junctions display significantly shorter $\xi_{F1}$ and $\xi_{F2}$ characteristic lengthscales. We attribute the short $\xi_{F1}$ to the common property of amorphous alloys having a short $l_e$ due to structural disorder, however scattering at the Pt/Co$_{68}$B$_{32}$ interfaces may also be considerable for very thin \textit{F} layers. The advantage of the short $\xi_{F2}$ in our system is that we have access to the first zero-$\pi$ and $\pi$-zero transitions before the \textit{F} layer undergoes the reorientation transition to in-plane magnetisation. Despite the short $\xi_{F1}$, the extrapolated $I_cR_N$ at zero thickness for our junctions, $V_0 = 56\pm8~\mu$V, is comparable to Ni$_{80}$Fe$_{20}$ junctions, $V_0 = 69\pm19~\mu$V \cite{doi:10.1063/1.4989392}, which have been extensively studied for similar applications \cite{PhysRevB.71.180501,PhysRevLett.97.177003,PhysRevB.76.094522,4277679,doi:10.1063/1.4862195,doi:10.1063/1.4989392,PhysRevB.97.024517, bell_controllable_2004,gingrich2016controllable,dayton2017experimental,Madden_2018}. In addition, the $I_cR_N$ product at the peak of the first $\pi$ state, $V_\pi \approx 7 \mu$V, is comparable to other ferromagnetic alloys, for example $V_\pi \approx 5 \mu$V in Ni$_{65}$Fe$_{15}$Co$_{20}$ and  $V_\pi \approx 12 \mu$V in Ni$_{80}$Fe$_{20}$ \cite{doi:10.1063/1.4989392}. The disadvantage of the short $\xi_{F2}$ for application is that precise control over thickness is necessary, since small variations will cause large changes to the $I_cR_N$ product and could potentially change the phase difference across the junction. Fortunately, such PMA multilayer stacks have an established industrial process for applications in magnetic recording.

The key application of ferromagnetic Josephson junctions is Josephson magnetic RAM (JMRAM) \cite{herr2012josephson}. In this technology, the Co$_{68}$B$_{32}$ forms one layer in a pseudospin-valve Josephson junction of the general type reported in our previous publication where we used Co and Co$_{68}$B$_{32}$ \cite{SatchellPSV}. In order to use the pseudospin-valve for JMRAM, the component ferromagnetic layers must be well characterised, as we report here for Pt/Co$_{68}$B$_{32}$/Pt PMA trilayers. In JMRAM, the zero-$\pi$ ferromagnetic junction is a passive phase shifter in a SQUID loop containing two $S/I/S$ junctions. As the zero-$\pi$ junction is passive, there is no need for this junction to have a large $I_cR_N$, however as we demonstrate our Co$_{68}$B$_{32}$ has comparable performance in this regard to Ni$_{80}$Fe$_{20}$, which is used in state-of-the-art devices. A JMRAM memory cell with in-plane ferromagnets (Ni$_{80}$Fe$_{20}$ and Ni) was recently demonstrated by Dayton \textit{et al.} \cite{dayton2017experimental}.

The detrimental features of magnetisation in-plane junctions are driven by the interaction between the physics underlying the Fraunhofer response and the vector potential of the ferromagnet. For in-plane single domain junctions, the Fraunhofer pattern will be uniformly shifted from zero global applied field by the vector potential of the ferromagnet in the plane of the junction. For example, Glick \textit{et al.} characterise candidate in-plane $F$ layers\cite{doi:10.1063/1.4989392}, where the Fraunhofer patterns are uniformly shifted by $H_\text{shift}\propto -Md_F$. To ensure that the $F$ layers are single domain, the maximum junction area in that work is $0.5\mu$m$^2$. If the area of the junctions is increased so that the ferromagnet becomes multidomain, it may not be possible to recover a Fraunhofer pattern due to the distortion by the stray fields emanating from the domains\cite{Bourgeois,PhysRevB.79.094523,PhysRevB.80.020506}. Such size considerations place an upper limit on the critical current of devices. Finally, the in-plane measurement field required to characterise the Fraunhofer pattern can cause premature switching of in-plane ferromagnets. Combined, the shift and premature switching means that JMRAM devices with in-plane ferromagnets may not have access to the highest critical current state, as observed for in-plane pseudospin-valve devices by Niedzielski \textit{et al.} \cite{niedzielski2017spin}.

In contrast, the use of PMA ferromagnets as we have demonstrated offers solutions to the issues introduced by in-plane ferromagnets. As the direction of the magnetisation and stray fields of PMA layers are parallel to the direction of current in the junction, they do not shift or distort the Fraunhofer pattern. Furthermore, due to the favourable demagnetisation effects, PMA materials systems have less stray fields and larger domain size limits compared to in-plane systems. Combined, these advantages allow us to successfully measure well defined Fraunhofer patterns on junctions with an area of $\approx 7\mu$m$^2$, much larger than those Glick \textit{et al.} require to characterise in-plane layers. As a result of the PMA, the highest critical current state of the junction is available at zero global applied field, as demonstrated in Figure \ref{IcB} (b). Finally, the application of in-plane measurement fields will not switch a PMA ferromagnet.

For practicable memory devices, it will be important to remove the need for an applied out-of-plane switching field. Such a field to switch a PMA ferromagnet may introduce flux vortices into the superconducting Nb electrodes, which we negated in this work by applying such fields only above $T_c$. Routes towards removing this restriction by implementing an all electrical switching process include spin-transfer torque and spin-orbit torque switching of the barriers \cite{miron2010current,miron2011perpendicular,PhysRevApplied.3.011001}.

\section*{Conclusions}

 In conclusion, we demonstrate Josephson $\pi$-junctions with Pt/Co$_{68}$B$_{32}$/Pt perpendicular magnetic anisotropy barriers. Co$_{68}$B$_{32}$ is a strong ferromagnetic amorphous alloy of interest in spintronics due to its low pinning properties. We show that at the Pt/Co$_{68}$B$_{32}$ interfaces there is significant polarisation of the Pt and that the samples are magnetic down to a nominal Co$_{68}$B$_{32}$ thickness of 0.2 nm. In Josephson junctions, as the thickness of Co$_{68}$B$_{32}$ is increased, we observe the nonmonotonic decay and oscillation of the critical Josephson current. These oscillations are attributed to the junctions undergoing the zero to $\pi$ transition. $\pi$-junctions have important applications in superconducting electronics, including cryogenic memory. Systematic material studies are crucial for the development of such technologies. The performance of our perpendicular magnetic anisotropy $\pi$-junctions are at least comparable to that of NiFe, which has in-plane magnetisation.

\section*{Methods}

Samples are deposited, fabricated, and measured using identical methodology to our previous work \cite{SatchellPSV}. The final product of cleanroom processing are standard ``sandwich'' planar Josephson junctions, defined by photolithography and Ar$^+$ ion milling, where the current flows perpendicular to the plane. The diameter of the circular junctions is a design parameter and is nominally $3~\mu$m. 

We dc sputter deposit the multilayer samples onto thermally oxidised Si substrates in the Royce Deposition System \cite{Royce}. The magnetrons are mounted below, and confocal to, the substrate with source-substrate distances of 134~mm. The base pressure of the vacuum chamber is 1$\times$10$^{-9}\units{mbar}$. The samples are deposited at room temperature with an Ar (6N purity) gas pressure of 3.6$\times$10$^{-3}\units{mbar}$ for the [Nb/Au]$_\text{x3}$/Nb bottom electrode layers and 4.8$\times$10$^{-3}\units{mbar}$ for the Pt/Co$_{68}$B$_{32}$/Pt barrier layers. The [Nb/Au]$_\text{x3}$/Nb superlattice is used for the bottom electrode as the superlattice has a lower surface roughness compared to a single Nb layer of comparable total thickness \cite{PhysRevB.85.214522,quarterman2020distortions}. Finally, a Nb/Au cap is deposited to prevent oxidation during the processing. In the final stage of sample fabrication, the top electrode, 150~nm of Nb, is deposited after an \textit{in-situ} ion milling process to remove 5~nm from the 10~nm Au cap. The full structure of the final device with thickness in (nm) is [Nb (25)/Au(2.5)]$_\text{x3}$/Nb (20)/Pt (10)/Co$_{68}$B$_{32}$ (\textit{d}$_\text{CoB}$)/ Pt (5)/Nb (5)/Au (5)/ Nb (150).

The choice of Pt thicknesses in this study is informed by previous results \cite{PhysRevB.99.174519}, being a balance between developing a suitable textured surface on which to grow the Co$_{68}$B$_{32}$ layer whilst not being in a range that significantly affects the critical current density. The total thickness of the Pt in this work is the same as that required for pseudospin-valve devices \cite{SatchellPSV}.

Fabricated devices are measured in a continuous flow $^4\mathrm{He}$ cryostat with 3\units{T} horizontal superconducting Helmholtz coils. The sample can be rotated about the vertical axis, which we perform in increments of 90$^{\circ}$ to bring that field in- and out-of-plane of the junctions. In order to avoid trapping flux in the superconducting Nb layers in the devices, we performed all sample rotations in zero field above the $T_c$ of the Nb and always cooled the sample in zero applied field (in practice there will inevitably be a small remanent field due to trapped flux in the magnet). The full sequence of setting the magnetic state of our samples and performing the measurement field sweeps is: warm to 15K, rotate sample to apply field out-of-plane, apply saturating field, remove saturating field, rotate sample by 90$^{\circ}$, cool sample, apply field in-plane, measure. Once we have finished measuring that magnetic state of the sample, we may wish to measure a further condition, such as reversing the magnetisation. To do so, we remove the in-plane field, warm to 15K and repeat the cycle described above.

Traditional 4-point-probe transport geometry is used to measure the current-voltage characteristic of the junction with combined Keithley 6221-2182A current source and nano-voltmeter. Magnetisation loops of sheet films are measured using a Quantum Design MPMS 3 magnetometer.

\section*{Data Availability}

The datasets generated during the current study are available in the University of Leeds repository, https://doi.org/10.5518/817.

\bibliography{library}

\begin{thebibliography}{10}
\urlstyle{rm}
\expandafter\ifx\csname url\endcsname\relax
  \def\url#1{\texttt{#1}}\fi
\expandafter\ifx\csname urlprefix\endcsname\relax\def\urlprefix{URL }\fi
\expandafter\ifx\csname doiprefix\endcsname\relax\def\doiprefix{DOI: }\fi
\providecommand{\bibinfo}[2]{#2}
\providecommand{\eprint}[2][]{\url{#2}}

\bibitem{RevModPhys.77.935}
\bibinfo{author}{Buzdin, A.~I.}
\newblock \bibinfo{journal}{\bibinfo{title}{Proximity effects in
  superconductor-ferromagnet heterostructures}}.
\newblock {\emph{\JournalTitle{Rev. Mod. Phys.}}}
  \textbf{\bibinfo{volume}{77}}, \bibinfo{pages}{935--976},
  \doiprefix\url{10.1103/RevModPhys.77.935} (\bibinfo{year}{2005}).

\bibitem{RevModPhys.77.1321}
\bibinfo{author}{Bergeret, F.~S.}, \bibinfo{author}{Volkov, A.~F.} \&
  \bibinfo{author}{Efetov, K.~B.}
\newblock \bibinfo{journal}{\bibinfo{title}{Odd triplet superconductivity and
  related phenomena in superconductor-ferromagnet structures}}.
\newblock {\emph{\JournalTitle{Rev. Mod. Phys.}}}
  \textbf{\bibinfo{volume}{77}}, \bibinfo{pages}{1321--1373},
  \doiprefix\url{10.1103/RevModPhys.77.1321} (\bibinfo{year}{2005}).

\bibitem{eschrig_spin-polarized_2011}
\bibinfo{author}{Eschrig, M.}
\newblock \bibinfo{journal}{\bibinfo{title}{Spin-polarized supercurrents for
  spintronics}}.
\newblock {\emph{\JournalTitle{Phys. Today}}} \textbf{\bibinfo{volume}{64}},
  \bibinfo{pages}{43--49}, \doiprefix\url{10.1063/1.3541944}
  (\bibinfo{year}{2011}).

\bibitem{linder_superconducting_2015}
\bibinfo{author}{Linder, J.} \& \bibinfo{author}{Robinson, J. W.~A.}
\newblock \bibinfo{journal}{\bibinfo{title}{Superconducting spintronics}}.
\newblock {\emph{\JournalTitle{Nat. Phys.}}} \textbf{\bibinfo{volume}{11}},
  \bibinfo{pages}{307--315}, \doiprefix\url{10.1038/nphys3242}
  (\bibinfo{year}{2015}).

\bibitem{0034-4885-78-10-104501}
\bibinfo{author}{Eschrig, M.}
\newblock \bibinfo{journal}{\bibinfo{title}{Spin-polarized supercurrents for
  spintronics: a review of current progress}}.
\newblock {\emph{\JournalTitle{Rep. Prog. Phys}}}
  \textbf{\bibinfo{volume}{78}}, \bibinfo{pages}{104501},
  \doiprefix\url{10.1088/0034-4885/78/10/104501} (\bibinfo{year}{2015}).

\bibitem{doi:10.1098/rsta.2015.0150}
\bibinfo{author}{Birge, N.~O.}
\newblock \bibinfo{journal}{\bibinfo{title}{Spin-triplet supercurrents in
  {J}osephson junctions containing strong ferromagnetic materials}}.
\newblock {\emph{\JournalTitle{Philos. Trans. Royal Soc. A}}}
  \textbf{\bibinfo{volume}{376}}, \bibinfo{pages}{20150150},
  \doiprefix\url{10.1098/rsta.2015.0150} (\bibinfo{year}{2018}).

\bibitem{buzdin1982critical}
\bibinfo{author}{Buzdin, A.~I.}, \bibinfo{author}{Bulaevskii, L.~N.} \&
  \bibinfo{author}{Panyukov, S.~V.}
\newblock \bibinfo{journal}{\bibinfo{title}{Critical-current oscillations as a
  function of the exchange field and thickness of the ferromagnetic metal ({F})
  in an {SFS} {J}osephson junction}}.
\newblock {\emph{\JournalTitle{JETP Lett}}} \textbf{\bibinfo{volume}{35}},
  \bibinfo{pages}{178--180} (\bibinfo{year}{1982}).

\bibitem{PhysRevLett.86.2427}
\bibinfo{author}{Ryazanov, V.~V.} \emph{et~al.}
\newblock \bibinfo{journal}{\bibinfo{title}{Coupling of {T}wo {S}uperconductors
  through a {F}erromagnet: {E}vidence for a $\ensuremath{\pi}$ {J}unction}}.
\newblock {\emph{\JournalTitle{Phys. Rev. Lett.}}}
  \textbf{\bibinfo{volume}{86}}, \bibinfo{pages}{2427--2430},
  \doiprefix\url{10.1103/PhysRevLett.86.2427} (\bibinfo{year}{2001}).

\bibitem{PhysRevB.68.054531}
\bibinfo{author}{Sellier, H.}, \bibinfo{author}{Baraduc, C.},
  \bibinfo{author}{Lefloch, F.} \& \bibinfo{author}{Calemczuk, R.}
\newblock \bibinfo{journal}{\bibinfo{title}{Temperature-induced crossover
  between $0$ and $\ensuremath{\pi}$ states in {S/F/S} junctions}}.
\newblock {\emph{\JournalTitle{Phys. Rev. B}}} \textbf{\bibinfo{volume}{68}},
  \bibinfo{pages}{054531}, \doiprefix\url{10.1103/PhysRevB.68.054531}
  (\bibinfo{year}{2003}).

\bibitem{PhysRevLett.96.197003}
\bibinfo{author}{Oboznov, V.~A.}, \bibinfo{author}{Bol'ginov, V.~V.},
  \bibinfo{author}{Feofanov, A.~K.}, \bibinfo{author}{Ryazanov, V.~V.} \&
  \bibinfo{author}{Buzdin, A.~I.}
\newblock \bibinfo{journal}{\bibinfo{title}{{Thickness Dependence of the
  Josephson Ground States of Superconductor-Ferromagnet-Superconductor
  Junctions}}}.
\newblock {\emph{\JournalTitle{Phys. Rev. Lett.}}}
  \textbf{\bibinfo{volume}{96}}, \bibinfo{pages}{197003},
  \doiprefix\url{10.1103/PhysRevLett.96.197003} (\bibinfo{year}{2006}).

\bibitem{PhysRevLett.97.247001}
\bibinfo{author}{Weides, M.} \emph{et~al.}
\newblock
  \bibinfo{journal}{\bibinfo{title}{$0\mathrm{\text{\ensuremath{-}}}\ensuremath{\pi}$
  {Josephson Tunnel Junctions with Ferromagnetic Barrier}}}.
\newblock {\emph{\JournalTitle{Phys. Rev. Lett.}}}
  \textbf{\bibinfo{volume}{97}}, \bibinfo{pages}{247001},
  \doiprefix\url{10.1103/PhysRevLett.97.247001} (\bibinfo{year}{2006}).

\bibitem{doi:10.1063/1.2356104}
\bibinfo{author}{Weides, M.} \emph{et~al.}
\newblock \bibinfo{journal}{\bibinfo{title}{High quality ferromagnetic 0 and
  $\pi$ {J}osephson tunnel junctions}}.
\newblock {\emph{\JournalTitle{Appl. Phys. Lett.}}}
  \textbf{\bibinfo{volume}{89}}, \bibinfo{pages}{122511},
  \doiprefix\url{10.1063/1.2356104} (\bibinfo{year}{2006}).

\bibitem{PhysRevLett.121.177702}
\bibinfo{author}{Stoutimore, M. J.~A.} \emph{et~al.}
\newblock \bibinfo{journal}{\bibinfo{title}{{Second-Harmonic Current-Phase
  Relation in Josephson Junctions with Ferromagnetic Barriers}}}.
\newblock {\emph{\JournalTitle{Phys. Rev. Lett.}}}
  \textbf{\bibinfo{volume}{121}}, \bibinfo{pages}{177702},
  \doiprefix\url{10.1103/PhysRevLett.121.177702} (\bibinfo{year}{2018}).

\bibitem{bolginov2018fabrication}
\bibinfo{author}{Bolginov, V.~V.}, \bibinfo{author}{Rossolenko, A.~N.},
  \bibinfo{author}{Shkarin, A.~B.}, \bibinfo{author}{Oboznov, V.~A.} \&
  \bibinfo{author}{Ryazanov, V.~V.}
\newblock \bibinfo{journal}{\bibinfo{title}{{Fabrication of Optimized
  Superconducting Phase Inverters Based on
  Superconductor--Ferromagnet--Superconductor} $\pi$-{J}unctions}}.
\newblock {\emph{\JournalTitle{J. Low Temp. Phys.}}}
  \textbf{\bibinfo{volume}{190}}, \bibinfo{pages}{302--314},
  \doiprefix\url{10.1007/s10909-017-1843-6} (\bibinfo{year}{2018}).

\bibitem{PhysRevLett.89.137007}
\bibinfo{author}{Kontos, T.} \emph{et~al.}
\newblock \bibinfo{journal}{\bibinfo{title}{Josephson {J}unction through a
  {T}hin {F}erromagnetic {L}ayer: {N}egative {C}oupling}}.
\newblock {\emph{\JournalTitle{Phys. Rev. Lett.}}}
  \textbf{\bibinfo{volume}{89}}, \bibinfo{pages}{137007},
  \doiprefix\url{10.1103/PhysRevLett.89.137007} (\bibinfo{year}{2002}).

\bibitem{PhysRevB.79.094523}
\bibinfo{author}{Khaire, T.~S.}, \bibinfo{author}{Pratt, W.~P.} \&
  \bibinfo{author}{Birge, N.~O.}
\newblock \bibinfo{journal}{\bibinfo{title}{Critical current behavior in
  {J}osephson junctions with the weak ferromagnet {PdNi}}}.
\newblock {\emph{\JournalTitle{Phys. Rev. B}}} \textbf{\bibinfo{volume}{79}},
  \bibinfo{pages}{094523}, \doiprefix\url{10.1103/PhysRevB.79.094523}
  (\bibinfo{year}{2009}).

\bibitem{7747519}
\bibinfo{author}{{Glick}, J.~A.}, \bibinfo{author}{{Loloee}, R.},
  \bibinfo{author}{{Pratt}, W.~P.} \& \bibinfo{author}{{Birge}, N.~O.}
\newblock \bibinfo{journal}{\bibinfo{title}{{Critical Current Oscillations of
  Josephson Junctions Containing PdFe Nanomagnets}}}.
\newblock {\emph{\JournalTitle{IEEE Trans. Appl. Supercond.}}}
  \textbf{\bibinfo{volume}{27}}, \bibinfo{pages}{1--5},
  \doiprefix\url{10.1109/TASC.2016.2630024} (\bibinfo{year}{2017}).

\bibitem{PhysRevLett.89.187004}
\bibinfo{author}{Blum, Y.}, \bibinfo{author}{Tsukernik, A.},
  \bibinfo{author}{Karpovski, M.} \& \bibinfo{author}{Palevski, A.}
\newblock \bibinfo{journal}{\bibinfo{title}{{Oscillations of the
  Superconducting Critical Current in Nb-Cu-Ni-Cu-Nb Junctions}}}.
\newblock {\emph{\JournalTitle{Phys. Rev. Lett.}}}
  \textbf{\bibinfo{volume}{89}}, \bibinfo{pages}{187004},
  \doiprefix\url{10.1103/PhysRevLett.89.187004} (\bibinfo{year}{2002}).

\bibitem{PhysRevB.73.174506}
\bibinfo{author}{Shelukhin, V.} \emph{et~al.}
\newblock \bibinfo{journal}{\bibinfo{title}{Observation of periodic
  $\ensuremath{\pi}$-phase shifts in ferromagnet-superconductor multilayers}}.
\newblock {\emph{\JournalTitle{Phys. Rev. B}}} \textbf{\bibinfo{volume}{73}},
  \bibinfo{pages}{174506}, \doiprefix\url{10.1103/PhysRevB.73.174506}
  (\bibinfo{year}{2006}).

\bibitem{PhysRevLett.97.177003}
\bibinfo{author}{Robinson, J. W.~A.}, \bibinfo{author}{Piano, S.},
  \bibinfo{author}{Burnell, G.}, \bibinfo{author}{Bell, C.} \&
  \bibinfo{author}{Blamire, M.~G.}
\newblock \bibinfo{journal}{\bibinfo{title}{Critical {Current Oscillations in
  Strong F}erromagnetic $\ensuremath{\pi}$ {J}unctions}}.
\newblock {\emph{\JournalTitle{Phys. Rev. Lett.}}}
  \textbf{\bibinfo{volume}{97}}, \bibinfo{pages}{177003},
  \doiprefix\url{10.1103/PhysRevLett.97.177003} (\bibinfo{year}{2006}).

\bibitem{PhysRevB.76.094522}
\bibinfo{author}{Robinson, J. W.~A.}, \bibinfo{author}{Piano, S.},
  \bibinfo{author}{Burnell, G.}, \bibinfo{author}{Bell, C.} \&
  \bibinfo{author}{Blamire, M.~G.}
\newblock \bibinfo{journal}{\bibinfo{title}{Zero to $\ensuremath{\pi}$
  transition in superconductor-ferromagnet-superconductor junctions}}.
\newblock {\emph{\JournalTitle{Phys. Rev. B}}} \textbf{\bibinfo{volume}{76}},
  \bibinfo{pages}{094522}, \doiprefix\url{10.1103/PhysRevB.76.094522}
  (\bibinfo{year}{2007}).

\bibitem{4277679}
\bibinfo{author}{{Robinson}, J. W.~A.}, \bibinfo{author}{{Piano}, S.},
  \bibinfo{author}{{Burnell}, G.}, \bibinfo{author}{{Bell}, C.} \&
  \bibinfo{author}{{Blamire}, M.~G.}
\newblock \bibinfo{journal}{\bibinfo{title}{Transport and {M}agnetic
  {P}roperties of {S}trong {F}erromagnetic {P}i-{J}unctions}}.
\newblock {\emph{\JournalTitle{IEEE Trans. Appl. Supercond.}}}
  \textbf{\bibinfo{volume}{17}}, \bibinfo{pages}{641--644},
  \doiprefix\url{10.1109/TASC.2007.898720} (\bibinfo{year}{2007}).

\bibitem{PhysRevB.79.054501}
\bibinfo{author}{Bannykh, A.~A.} \emph{et~al.}
\newblock \bibinfo{journal}{\bibinfo{title}{Josephson tunnel junctions with a
  strong ferromagnetic interlayer}}.
\newblock {\emph{\JournalTitle{Phys. Rev. B}}} \textbf{\bibinfo{volume}{79}},
  \bibinfo{pages}{054501}, \doiprefix\url{10.1103/PhysRevB.79.054501}
  (\bibinfo{year}{2009}).

\bibitem{baek2014hybrid}
\bibinfo{author}{Baek, B.}, \bibinfo{author}{Rippard, W.~H.},
  \bibinfo{author}{Benz, S.~P.}, \bibinfo{author}{Russek, S.~E.} \&
  \bibinfo{author}{Dresselhaus, P.~D.}
\newblock \bibinfo{journal}{\bibinfo{title}{Hybrid superconducting-magnetic
  memory device using competing order parameters}}.
\newblock {\emph{\JournalTitle{Nat. Commun.}}} \textbf{\bibinfo{volume}{5}},
  \bibinfo{pages}{3888}, \doiprefix\url{10.1038/ncomms4888}
  (\bibinfo{year}{2014}).

\bibitem{PhysRevApplied.7.064013}
\bibinfo{author}{Baek, B.}, \bibinfo{author}{Schneider, M.~L.},
  \bibinfo{author}{Pufall, M.~R.} \& \bibinfo{author}{Rippard, W.~H.}
\newblock \bibinfo{journal}{\bibinfo{title}{{Phase Offsets in the
  Critical-Current Oscillations of Josephson Junctions Based on Ni and
  Ni}-($\text{Ni}_{81}\text{Fe}_{19})_{x}\text{Nb}_{y}$ {B}arriers}}.
\newblock {\emph{\JournalTitle{Phys. Rev. Applied}}}
  \textbf{\bibinfo{volume}{7}}, \bibinfo{pages}{064013},
  \doiprefix\url{10.1103/PhysRevApplied.7.064013} (\bibinfo{year}{2017}).

\bibitem{8359359}
\bibinfo{author}{{Baek}, B.}, \bibinfo{author}{{Schneider}, M.~L.},
  \bibinfo{author}{{Pufall}, M.~R.} \& \bibinfo{author}{{Rippard}, W.~H.}
\newblock \bibinfo{journal}{\bibinfo{title}{{Anomalous Supercurrent Modulation
  in Josephson Junctions With Ni-Based Barriers}}}.
\newblock {\emph{\JournalTitle{IEEE Trans. Appl. Supercond.}}}
  \textbf{\bibinfo{volume}{28}}, \bibinfo{pages}{1--5},
  \doiprefix\url{10.1109/TASC.2018.2836961} (\bibinfo{year}{2018}).

\bibitem{doi:10.1063/1.3262969}
\bibinfo{author}{Robinson, J. W.~A.}, \bibinfo{author}{Barber, Z.~H.} \&
  \bibinfo{author}{Blamire, M.~G.}
\newblock \bibinfo{journal}{\bibinfo{title}{Strong ferromagnetic {J}osephson
  devices with optimized magnetism}}.
\newblock {\emph{\JournalTitle{Appl. Phys. Lett.}}}
  \textbf{\bibinfo{volume}{95}}, \bibinfo{pages}{192509},
  \doiprefix\url{10.1063/1.3262969} (\bibinfo{year}{2009}).

\bibitem{piano20070}
\bibinfo{author}{Piano, S.}, \bibinfo{author}{Robinson, J. W.~A.},
  \bibinfo{author}{Burnell, G.} \& \bibinfo{author}{Blamire, M.~G.}
\newblock \bibinfo{journal}{\bibinfo{title}{0-$\pi$ oscillations in
  nanostructured {Nb/Fe/Nb J}osephson junctions}}.
\newblock {\emph{\JournalTitle{Eur. Phys. J. B}}}
  \textbf{\bibinfo{volume}{58}}, \bibinfo{pages}{123--126},
  \doiprefix\url{10.1140/epjb/e2007-00210-8} (\bibinfo{year}{2007}).

\bibitem{PhysRevB.71.180501}
\bibinfo{author}{Bell, C.}, \bibinfo{author}{Loloee, R.},
  \bibinfo{author}{Burnell, G.} \& \bibinfo{author}{Blamire, M.~G.}
\newblock \bibinfo{journal}{\bibinfo{title}{Characteristics of strong
  ferromagnetic {J}osephson junctions with epitaxial barriers}}.
\newblock {\emph{\JournalTitle{Phys. Rev. B}}} \textbf{\bibinfo{volume}{71}},
  \bibinfo{pages}{180501(R)}, \doiprefix\url{10.1103/PhysRevB.71.180501}
  (\bibinfo{year}{2005}).

\bibitem{doi:10.1063/1.4862195}
\bibinfo{author}{Abd El~Qader, M.} \emph{et~al.}
\newblock \bibinfo{journal}{\bibinfo{title}{Switching at small magnetic fields
  in {J}osephson junctions fabricated with ferromagnetic barrier layers}}.
\newblock {\emph{\JournalTitle{Appl. Phys. Lett.}}}
  \textbf{\bibinfo{volume}{104}}, \bibinfo{pages}{022602},
  \doiprefix\url{10.1063/1.4862195} (\bibinfo{year}{2014}).

\bibitem{doi:10.1063/1.4989392}
\bibinfo{author}{Glick, J.~A.} \emph{et~al.}
\newblock \bibinfo{journal}{\bibinfo{title}{Critical current oscillations of
  elliptical {J}osephson junctions with single-domain ferromagnetic layers}}.
\newblock {\emph{\JournalTitle{J. Appl. Phys.}}}
  \textbf{\bibinfo{volume}{122}}, \bibinfo{pages}{133906},
  \doiprefix\url{10.1063/1.4989392} (\bibinfo{year}{2017}).

\bibitem{PhysRevB.97.024517}
\bibinfo{author}{Niedzielski, B.~M.} \emph{et~al.}
\newblock \bibinfo{journal}{\bibinfo{title}{Spin-valve {J}osephson junctions
  for cryogenic memory}}.
\newblock {\emph{\JournalTitle{Phys. Rev. B}}} \textbf{\bibinfo{volume}{97}},
  \bibinfo{pages}{024517}, \doiprefix\url{10.1103/PhysRevB.97.024517}
  (\bibinfo{year}{2018}).

\bibitem{PhysRevB.74.140501}
\bibinfo{author}{Born, F.} \emph{et~al.}
\newblock \bibinfo{journal}{\bibinfo{title}{Multiple
  $0\text{\ensuremath{-}}\ensuremath{\pi}$ transitions in
  superconductor/insulator/ferromagnet/superconductor {J}osephson tunnel
  junctions}}.
\newblock {\emph{\JournalTitle{Phys. Rev. B}}} \textbf{\bibinfo{volume}{74}},
  \bibinfo{pages}{140501(R)}, \doiprefix\url{10.1103/PhysRevB.74.140501}
  (\bibinfo{year}{2006}).

\bibitem{Niedzielski_2015}
\bibinfo{author}{Niedzielski, B.~M.}, \bibinfo{author}{Gingrich, E.~C.},
  \bibinfo{author}{Loloee, R.}, \bibinfo{author}{Pratt, W.~P.} \&
  \bibinfo{author}{Birge, N.~O.}
\newblock \bibinfo{journal}{\bibinfo{title}{{S/F/S J}osephson junctions with
  single-domain ferromagnets for memory applications}}.
\newblock {\emph{\JournalTitle{Supercond. Sci. Technol.}}}
  \textbf{\bibinfo{volume}{28}}, \bibinfo{pages}{085012},
  \doiprefix\url{10.1088/0953-2048/28/8/085012} (\bibinfo{year}{2015}).

\bibitem{Margaris_2010}
\bibinfo{author}{Margaris, I.}, \bibinfo{author}{Paltoglou, V.} \&
  \bibinfo{author}{Flytzanis, N.}
\newblock \bibinfo{journal}{\bibinfo{title}{Zero phase difference supercurrent
  in ferromagnetic {J}osephson junctions}}.
\newblock {\emph{\JournalTitle{J. Phys. Condens. Matter}}}
  \textbf{\bibinfo{volume}{22}}, \bibinfo{pages}{445701},
  \doiprefix\url{10.1088/0953-8984/22/44/445701} (\bibinfo{year}{2010}).

\bibitem{PhysRevB.92.174516}
\bibinfo{author}{Halterman, K.}, \bibinfo{author}{Valls, O.~T.} \&
  \bibinfo{author}{Wu, C.-T.}
\newblock \bibinfo{journal}{\bibinfo{title}{Charge and spin currents in
  ferromagnetic {J}osephson junctions}}.
\newblock {\emph{\JournalTitle{Phys. Rev. B}}} \textbf{\bibinfo{volume}{92}},
  \bibinfo{pages}{174516}, \doiprefix\url{10.1103/PhysRevB.92.174516}
  (\bibinfo{year}{2015}).

\bibitem{PhysRevB.95.184508}
\bibinfo{author}{Silaev, M.~A.}, \bibinfo{author}{Tokatly, I.~V.} \&
  \bibinfo{author}{Bergeret, F.~S.}
\newblock \bibinfo{journal}{\bibinfo{title}{Anomalous current in diffusive
  ferromagnetic {J}osephson junctions}}.
\newblock {\emph{\JournalTitle{Phys. Rev. B}}} \textbf{\bibinfo{volume}{95}},
  \bibinfo{pages}{184508}, \doiprefix\url{10.1103/PhysRevB.95.184508}
  (\bibinfo{year}{2017}).

\bibitem{PhysRevB.86.224506}
\bibinfo{author}{Gingrich, E.~C.} \emph{et~al.}
\newblock \bibinfo{journal}{\bibinfo{title}{Spin-triplet supercurrent in {Co/Ni
  multilayer J}osephson junctions with perpendicular anisotropy}}.
\newblock {\emph{\JournalTitle{Phys. Rev. B}}} \textbf{\bibinfo{volume}{86}},
  \bibinfo{pages}{224506}, \doiprefix\url{10.1103/PhysRevB.86.224506}
  (\bibinfo{year}{2012}).

\bibitem{PhysRevB.96.224515}
\bibinfo{author}{Glick, J.~A.} \emph{et~al.}
\newblock \bibinfo{journal}{\bibinfo{title}{Spin-triplet supercurrent in
  {J}osephson junctions containing a synthetic antiferromagnet with
  perpendicular magnetic anisotropy}}.
\newblock {\emph{\JournalTitle{Phys. Rev. B}}} \textbf{\bibinfo{volume}{96}},
  \bibinfo{pages}{224515}, \doiprefix\url{10.1103/PhysRevB.96.224515}
  (\bibinfo{year}{2017}).

\bibitem{Glickeaat9457}
\bibinfo{author}{Glick, J.~A.} \emph{et~al.}
\newblock \bibinfo{journal}{\bibinfo{title}{Phase control in a spin-triplet
  {SQUID}}}.
\newblock {\emph{\JournalTitle{Sci. Adv.}}} \textbf{\bibinfo{volume}{4}},
  \bibinfo{pages}{eaat9457}, \doiprefix\url{10.1126/sciadv.aat9457}
  (\bibinfo{year}{2018}).

\bibitem{PhysRevB.99.174519}
\bibinfo{author}{Satchell, N.}, \bibinfo{author}{Loloee, R.} \&
  \bibinfo{author}{Birge, N.~O.}
\newblock \bibinfo{journal}{\bibinfo{title}{Supercurrent in ferromagnetic
  {J}osephson junctions with heavy-metal interlayers. {II. C}anted
  magnetization}}.
\newblock {\emph{\JournalTitle{Phys. Rev. B}}} \textbf{\bibinfo{volume}{99}},
  \bibinfo{pages}{174519}, \doiprefix\url{10.1103/PhysRevB.99.174519}
  (\bibinfo{year}{2019}).

\bibitem{PhysRevB.47.2671}
\bibinfo{author}{Tanaka, H.} \emph{et~al.}
\newblock \bibinfo{journal}{\bibinfo{title}{Electronic structure and magnetism
  of amorphous {Co}$_{1-x}${B}$_{x}$ alloys}}.
\newblock {\emph{\JournalTitle{Phys. Rev. B}}} \textbf{\bibinfo{volume}{47}},
  \bibinfo{pages}{2671--2677}, \doiprefix\url{10.1103/PhysRevB.47.2671}
  (\bibinfo{year}{1993}).

\bibitem{doi:10.1063/1.3280373}
\bibinfo{author}{Lavrijsen, R.} \emph{et~al.}
\newblock \bibinfo{journal}{\bibinfo{title}{Reduced domain wall pinning in
  ultrathin {P}t/{C}o$_{100−x}${B}$_x$/{P}t with perpendicular magnetic
  anisotropy}}.
\newblock {\emph{\JournalTitle{Appl. Phys. Lett.}}}
  \textbf{\bibinfo{volume}{96}}, \bibinfo{pages}{022501},
  \doiprefix\url{10.1063/1.3280373} (\bibinfo{year}{2010}).

\bibitem{schellekens2012electric}
\bibinfo{author}{Schellekens, A.~J.}, \bibinfo{author}{Van~den Brink, A.},
  \bibinfo{author}{Franken, J.~H.}, \bibinfo{author}{Swagten, H. J.~M.} \&
  \bibinfo{author}{Koopmans, B.}
\newblock \bibinfo{journal}{\bibinfo{title}{Electric-field control of domain
  wall motion in perpendicularly magnetized materials}}.
\newblock {\emph{\JournalTitle{Nat. Commun.}}} \textbf{\bibinfo{volume}{3}},
  \bibinfo{pages}{847}, \doiprefix\url{10.1038/ncomms1848}
  (\bibinfo{year}{2012}).

\bibitem{finizio2019deterministic}
\bibinfo{author}{Finizio, S.} \emph{et~al.}
\newblock \bibinfo{journal}{\bibinfo{title}{{Deterministic Field-Free Skyrmion
  Nucleation at a Nanoengineered Injector D}evice}}.
\newblock {\emph{\JournalTitle{Nano Lett.}}} \textbf{\bibinfo{volume}{19}},
  \bibinfo{pages}{7246--7255}, \doiprefix\url{10.1021/acs.nanolett.9b02840}
  (\bibinfo{year}{2019}).

\bibitem{zeissler2020diameter}
\bibinfo{author}{Zeissler, K.} \emph{et~al.}
\newblock \bibinfo{journal}{\bibinfo{title}{Diameter-independent skyrmion
  {H}all angle observed in chiral magnetic multilayers}}.
\newblock {\emph{\JournalTitle{Nat. Commun.}}} \textbf{\bibinfo{volume}{11}},
  \bibinfo{pages}{428}, \doiprefix\url{10.1038/s41467-019-14232-9}
  (\bibinfo{year}{2020}).

\bibitem{SatchellPSV}
\bibinfo{author}{Satchell, N.} \emph{et~al.}
\newblock \bibinfo{journal}{\bibinfo{title}{Spin-valve {J}osephson junctions
  with perpendicular magnetic anisotropy for cryogenic memory}}.
\newblock {\emph{\JournalTitle{Appl. Phys. Lett.}}}
  \textbf{\bibinfo{volume}{116}}, \bibinfo{pages}{022601},
  \doiprefix\url{10.1063/1.5140095} (\bibinfo{year}{2020}).

\bibitem{bell_controllable_2004}
\bibinfo{author}{Bell, C.} \emph{et~al.}
\newblock \bibinfo{journal}{\bibinfo{title}{Controllable {Josephson} current
  through a pseudospin-valve structure}}.
\newblock {\emph{\JournalTitle{Appl. Phys. Lett.}}}
  \textbf{\bibinfo{volume}{84}}, \bibinfo{pages}{1153--1155},
  \doiprefix\url{http://dx.doi.org/10.1063/1.1646217} (\bibinfo{year}{2004}).

\bibitem{gingrich2016controllable}
\bibinfo{author}{Gingrich, E.~C.} \emph{et~al.}
\newblock \bibinfo{journal}{\bibinfo{title}{Controllable 0--$\pi$ {J}osephson
  junctions containing a ferromagnetic spin valve}}.
\newblock {\emph{\JournalTitle{Nat. Phys.}}} \textbf{\bibinfo{volume}{12}},
  \bibinfo{pages}{564}, \doiprefix\url{10.1038/nphys3681}
  (\bibinfo{year}{2016}).

\bibitem{dayton2017experimental}
\bibinfo{author}{Dayton, I.~M.} \emph{et~al.}
\newblock \bibinfo{journal}{\bibinfo{title}{Experimental demonstration of a
  {J}osephson magnetic memory cell with a programmable $\pi$-junction}}.
\newblock {\emph{\JournalTitle{IEEE Magn. Lett.}}}
  \textbf{\bibinfo{volume}{9}}, \bibinfo{pages}{3301905},
  \doiprefix\url{10.1109/LMAG.2018.2801820} (\bibinfo{year}{2018}).

\bibitem{Madden_2018}
\bibinfo{author}{Madden, A.~E.}, \bibinfo{author}{Willard, J.~C.},
  \bibinfo{author}{Loloee, R.} \& \bibinfo{author}{Birge, N.~O.}
\newblock \bibinfo{journal}{\bibinfo{title}{Phase controllable {J}osephson
  junctions for cryogenic memory}}.
\newblock {\emph{\JournalTitle{Supercond. Sci. Technol.}}}
  \textbf{\bibinfo{volume}{32}}, \bibinfo{pages}{015001},
  \doiprefix\url{10.1088/1361-6668/aae8cf} (\bibinfo{year}{2018}).

\bibitem{312324}
\bibinfo{author}{{Kon\u{c}}, M.} \emph{et~al.}
\newblock \bibinfo{journal}{\bibinfo{title}{Temperature dependence of the
  magnetization and of the other physical properties of rapidly quenched
  amorphous {CoB} alloys}}.
\newblock {\emph{\JournalTitle{IEEE Trans. Magn.}}}
  \textbf{\bibinfo{volume}{30}}, \bibinfo{pages}{524--526},
  \doiprefix\url{10.1109/20.312324} (\bibinfo{year}{1994}).

\bibitem{doi:10.1063/1.344903}
\bibinfo{author}{Schütz, G.} \emph{et~al.}
\newblock \bibinfo{journal}{\bibinfo{title}{Spin‐dependent x‐ray absorption
  in {Co/Pt} multilayers and {Co}$_{50}${P}t$_{50}$ alloy}}.
\newblock {\emph{\JournalTitle{J. Appl. Phys.}}} \textbf{\bibinfo{volume}{67}},
  \bibinfo{pages}{4456--4458}, \doiprefix\url{10.1063/1.344903}
  (\bibinfo{year}{1990}).

\bibitem{PhysRevB.65.020405}
\bibinfo{author}{Geissler, J.} \emph{et~al.}
\newblock \bibinfo{journal}{\bibinfo{title}{Pt magnetization profile in a
  {Pt/Co} bilayer studied by resonant magnetic x-ray reflectometry}}.
\newblock {\emph{\JournalTitle{Phys. Rev. B}}} \textbf{\bibinfo{volume}{65}},
  \bibinfo{pages}{020405(R)}, \doiprefix\url{10.1103/PhysRevB.65.020405}
  (\bibinfo{year}{2001}).

\bibitem{PhysRevB.72.054430}
\bibinfo{author}{Suzuki, M.} \emph{et~al.}
\newblock \bibinfo{journal}{\bibinfo{title}{Depth profile of spin and orbital
  magnetic moments in a subnanometer {P}t film on {C}o}}.
\newblock {\emph{\JournalTitle{Phys. Rev. B}}} \textbf{\bibinfo{volume}{72}},
  \bibinfo{pages}{054430}, \doiprefix\url{10.1103/PhysRevB.72.054430}
  (\bibinfo{year}{2005}).

\bibitem{rowan2017interfacial}
\bibinfo{author}{Rowan-Robinson, R.~M.} \emph{et~al.}
\newblock \bibinfo{journal}{\bibinfo{title}{The interfacial nature of
  proximity-induced magnetism and the {D}zyaloshinskii-{M}oriya interaction at
  the {Pt/Co} interface}}.
\newblock {\emph{\JournalTitle{Sci. Rep.}}} \textbf{\bibinfo{volume}{7}},
  \bibinfo{pages}{16835}, \doiprefix\url{10.1038/s41598-017-17137-z}
  (\bibinfo{year}{2017}).

\bibitem{PhysRevB.100.174418}
\bibinfo{author}{Inyang, O.} \emph{et~al.}
\newblock \bibinfo{journal}{\bibinfo{title}{Threshold interface magnetization
  required to induce magnetic proximity effect}}.
\newblock {\emph{\JournalTitle{Phys. Rev. B}}} \textbf{\bibinfo{volume}{100}},
  \bibinfo{pages}{174418}, \doiprefix\url{10.1103/PhysRevB.100.174418}
  (\bibinfo{year}{2019}).

\bibitem{PhysRevB.80.020506}
\bibinfo{author}{Khasawneh, M.~A.}, \bibinfo{author}{Pratt, W.~P.} \&
  \bibinfo{author}{Birge, N.~O.}
\newblock \bibinfo{journal}{\bibinfo{title}{Josephson junctions with a
  synthetic antiferromagnetic interlayer}}.
\newblock {\emph{\JournalTitle{Phys. Rev. B}}} \textbf{\bibinfo{volume}{80}},
  \bibinfo{pages}{020506(R)}, \doiprefix\url{10.1103/PhysRevB.80.020506}
  (\bibinfo{year}{2009}).

\bibitem{satchellSOC2018}
\bibinfo{author}{Satchell, N.} \& \bibinfo{author}{Birge, N.~O.}
\newblock \bibinfo{journal}{\bibinfo{title}{Supercurrent in ferromagnetic
  {J}osephson junctions with heavy metal interlayers}}.
\newblock {\emph{\JournalTitle{Phys. Rev. B}}} \textbf{\bibinfo{volume}{97}},
  \bibinfo{pages}{214509}, \doiprefix\url{10.1103/PhysRevB.97.214509}
  (\bibinfo{year}{2018}).

\bibitem{barone1982physics}
\bibinfo{author}{Barone, A.} \& \bibinfo{author}{Patern\`{o}, G.}
\newblock \emph{\bibinfo{title}{Physics and {A}pplications of the {J}osephson
  {E}ffect}} (\bibinfo{publisher}{John Wiley \& Sons, New York},
  \bibinfo{year}{1982}).

\bibitem{quarterman2020distortions}
\bibinfo{author}{Quarterman, P.} \emph{et~al.}
\newblock \bibinfo{journal}{\bibinfo{title}{Distortions to the penetration
  depth and coherence length of superconductor/normal-metal superlattices}}.
\newblock {\emph{\JournalTitle{Phys. Rev. Materials}}}
  \textbf{\bibinfo{volume}{4}}, \bibinfo{pages}{074801},
  \doiprefix\url{10.1103/PhysRevMaterials.4.074801} (\bibinfo{year}{2020}).

\bibitem{PhysRevLett.120.247001}
\bibinfo{author}{Flokstra, M.~G.} \emph{et~al.}
\newblock \bibinfo{journal}{\bibinfo{title}{Observation of {A}nomalous
  {M}eissner {S}creening in $\mathrm{Cu}/\mathrm{Nb}$ and
  $\mathrm{Cu}/\mathrm{Nb}/\mathrm{Co}$ {T}hin {F}ilms}}.
\newblock {\emph{\JournalTitle{Phys. Rev. Lett.}}}
  \textbf{\bibinfo{volume}{120}}, \bibinfo{pages}{247001},
  \doiprefix\url{10.1103/PhysRevLett.120.247001} (\bibinfo{year}{2018}).

\bibitem{PhysRevB.64.134506}
\bibinfo{author}{Bergeret, F.~S.}, \bibinfo{author}{Volkov, A.~F.} \&
  \bibinfo{author}{Efetov, K.~B.}
\newblock \bibinfo{journal}{\bibinfo{title}{Josephson current in
  superconductor-ferromagnet structures with a nonhomogeneous magnetization}}.
\newblock {\emph{\JournalTitle{Phys. Rev. B}}} \textbf{\bibinfo{volume}{64}},
  \bibinfo{pages}{134506}, \doiprefix\url{10.1103/PhysRevB.64.134506}
  (\bibinfo{year}{2001}).

\bibitem{PhysRevB.71.201403}
\bibinfo{author}{Krupin, O.} \emph{et~al.}
\newblock \bibinfo{journal}{\bibinfo{title}{Rashba effect at magnetic metal
  surfaces}}.
\newblock {\emph{\JournalTitle{Phys. Rev. B}}} \textbf{\bibinfo{volume}{71}},
  \bibinfo{pages}{201403}, \doiprefix\url{10.1103/PhysRevB.71.201403}
  (\bibinfo{year}{2005}).

\bibitem{miron2010current}
\bibinfo{author}{Miron, I.~M.} \emph{et~al.}
\newblock \bibinfo{journal}{\bibinfo{title}{Current-driven spin torque induced
  by the {R}ashba effect in a ferromagnetic metal layer}}.
\newblock {\emph{\JournalTitle{Nat. Mater.}}} \textbf{\bibinfo{volume}{9}},
  \bibinfo{pages}{230--234}, \doiprefix\url{10.1038/nmat2613}
  (\bibinfo{year}{2010}).

\bibitem{miron2011perpendicular}
\bibinfo{author}{Miron, I.~M.} \emph{et~al.}
\newblock \bibinfo{journal}{\bibinfo{title}{Perpendicular switching of a single
  ferromagnetic layer induced by in-plane current injection}}.
\newblock {\emph{\JournalTitle{Nature}}} \textbf{\bibinfo{volume}{476}},
  \bibinfo{pages}{189}, \doiprefix\url{10.1038/nature10309}
  (\bibinfo{year}{2011}).

\bibitem{PhysRevB.73.064505}
\bibinfo{author}{Faur\'e, M.}, \bibinfo{author}{Buzdin, A.~I.},
  \bibinfo{author}{Golubov, A.~A.} \& \bibinfo{author}{Kupriyanov, M.~Y.}
\newblock \bibinfo{journal}{\bibinfo{title}{Properties of
  superconductor/ferromagnet structures with spin-dependent scattering}}.
\newblock {\emph{\JournalTitle{Phys. Rev. B}}} \textbf{\bibinfo{volume}{73}},
  \bibinfo{pages}{064505}, \doiprefix\url{10.1103/PhysRevB.73.064505}
  (\bibinfo{year}{2006}).

\bibitem{PhysRevB.84.144513}
\bibinfo{author}{Pugach, N.~G.}, \bibinfo{author}{Kupriyanov, M.~Y.},
  \bibinfo{author}{Goldobin, E.}, \bibinfo{author}{Kleiner, R.} \&
  \bibinfo{author}{Koelle, D.}
\newblock
  \bibinfo{journal}{\bibinfo{title}{Superconductor-insulator-ferromagnet-superconductor
  {J}osephson junction: {F}rom the dirty to the clean limit}}.
\newblock {\emph{\JournalTitle{Phys. Rev. B}}} \textbf{\bibinfo{volume}{84}},
  \bibinfo{pages}{144513}, \doiprefix\url{10.1103/PhysRevB.84.144513}
  (\bibinfo{year}{2011}).

\bibitem{PhysRevB.89.134517}
\bibinfo{author}{Bergeret, F.~S.} \& \bibinfo{author}{Tokatly, I.~V.}
\newblock \bibinfo{journal}{\bibinfo{title}{Spin-orbit coupling as a source of
  long-range triplet proximity effect in superconductor-ferromagnet hybrid
  structures}}.
\newblock {\emph{\JournalTitle{Phys. Rev. B}}} \textbf{\bibinfo{volume}{89}},
  \bibinfo{pages}{134517}, \doiprefix\url{10.1103/PhysRevB.89.134517}
  (\bibinfo{year}{2014}).

\bibitem{jeon2018enhanced}
\bibinfo{author}{Jeon, K.-R.} \emph{et~al.}
\newblock \bibinfo{journal}{\bibinfo{title}{Enhanced spin pumping into
  superconductors provides evidence for superconducting pure spin currents}}.
\newblock {\emph{\JournalTitle{Nat. Mater.}}} \textbf{\bibinfo{volume}{17}},
  \bibinfo{pages}{499--503}, \doiprefix\url{10.1038/s41563-018-0058-9}
  (\bibinfo{year}{2018}).

\bibitem{herr2012josephson}
\bibinfo{author}{Herr, A.~Y.} \& \bibinfo{author}{Herr, Q.~P.}
\newblock \bibinfo{title}{Josephson magnetic random access memory system and
  method} (\bibinfo{year}{2012}).
\newblock \bibinfo{note}{{US Patent 8 270 209}}.

\bibitem{Bourgeois}
\bibinfo{author}{{Bourgeois, O.}} \emph{et~al.}
\newblock \bibinfo{journal}{\bibinfo{title}{Josephson effect through a
  ferromagnetic layer}}.
\newblock {\emph{\JournalTitle{Eur. Phys. J. B}}}
  \textbf{\bibinfo{volume}{21}}, \bibinfo{pages}{75--80},
  \doiprefix\url{10.1007/s100510170215} (\bibinfo{year}{2001}).

\bibitem{niedzielski2017spin}
\bibinfo{author}{Niedzielski, B.~M.} \emph{et~al.}
\newblock \bibinfo{journal}{\bibinfo{title}{Spin-valve {J}osephson junctions
  for cryogenic memory}}.
\newblock {\emph{\JournalTitle{Phys. Rev. B}}} \textbf{\bibinfo{volume}{97}},
  \bibinfo{pages}{024517}, \doiprefix\url{10.1103/PhysRevB.97.024517}
  (\bibinfo{year}{2018}).

\bibitem{PhysRevApplied.3.011001}
\bibinfo{author}{Baek, B.} \emph{et~al.}
\newblock \bibinfo{journal}{\bibinfo{title}{{Spin-Transfer Torque Switching in
  Nanopillar Superconducting-Magnetic Hybrid Josephson Junctions}}}.
\newblock {\emph{\JournalTitle{Phys. Rev. Applied}}}
  \textbf{\bibinfo{volume}{3}}, \bibinfo{pages}{011001},
  \doiprefix\url{10.1103/PhysRevApplied.3.011001} (\bibinfo{year}{2015}).

\bibitem{Royce}
\bibinfo{note}{The Royce Deposition System is a multi-chamber, multi-technique
  thin film deposition tool based at the University of Leeds as part of the
  \href{https://www.royce.ac.uk/}{Henry Royce Institute}.}

\bibitem{PhysRevB.85.214522}
\bibinfo{author}{Wang, Y.}, \bibinfo{author}{Pratt, W.~P.} \&
  \bibinfo{author}{Birge, N.~O.}
\newblock \bibinfo{journal}{\bibinfo{title}{Area-dependence of spin-triplet
  supercurrent in ferromagnetic {J}osephson junctions}}.
\newblock {\emph{\JournalTitle{Phys. Rev. B}}} \textbf{\bibinfo{volume}{85}},
  \bibinfo{pages}{214522}, \doiprefix\url{10.1103/PhysRevB.85.214522}
  (\bibinfo{year}{2012}).

\end{thebibliography}


\begin{thebibliography}{6}%
\makeatletter
\providecommand \@ifxundefined [1]{%
 \@ifx{#1\undefined}
}%
\providecommand \@ifnum [1]{%
 \ifnum #1\expandafter \@firstoftwo
 \else \expandafter \@secondoftwo
 \fi
}%
\providecommand \@ifx [1]{%
 \ifx #1\expandafter \@firstoftwo
 \else \expandafter \@secondoftwo
 \fi
}%
\providecommand \natexlab [1]{#1}%
\providecommand \enquote  [1]{``#1''}%
\providecommand \bibnamefont  [1]{#1}%
\providecommand \bibfnamefont [1]{#1}%
\providecommand \citenamefont [1]{#1}%
\providecommand \href@noop [0]{\@secondoftwo}%
\providecommand \href [0]{\begingroup \@sanitize@url \@href}%
\providecommand \@href[1]{\@@startlink{#1}\@@href}%
\providecommand \@@href[1]{\endgroup#1\@@endlink}%
\providecommand \@sanitize@url [0]{\catcode `\\12\catcode `\$12\catcode
  `\&12\catcode `\#12\catcode `\^12\catcode `\_12\catcode `\%12\relax}%
\providecommand \@@startlink[1]{}%
\providecommand \@@endlink[0]{}%
\providecommand \url  [0]{\begingroup\@sanitize@url \@url }%
\providecommand \@url [1]{\endgroup\@href {#1}{\urlprefix }}%
\providecommand \urlprefix  [0]{URL }%
\providecommand \Eprint [0]{\href }%
\providecommand \doibase [0]{http://dx.doi.org/}%
\providecommand \selectlanguage [0]{\@gobble}%
\providecommand \bibinfo  [0]{\@secondoftwo}%
\providecommand \bibfield  [0]{\@secondoftwo}%
\providecommand \translation [1]{[#1]}%
\providecommand \BibitemOpen [0]{}%
\providecommand \bibitemStop [0]{}%
\providecommand \bibitemNoStop [0]{.\EOS\space}%
\providecommand \EOS [0]{\spacefactor3000\relax}%
\providecommand \BibitemShut  [1]{\csname bibitem#1\endcsname}%
\let\auto@bib@innerbib\@empty
\bibitem [{\citenamefont {Garcia}\ \emph {et~al.}(2009)\citenamefont {Garcia},
  \citenamefont {Fernandez~Pinel}, \citenamefont {de~la Venta}, \citenamefont
  {Quesada}, \citenamefont {Bouzas}, \citenamefont {Fernández}, \citenamefont
  {Romero}, \citenamefont {Martín~González},\ and\ \citenamefont
  {Costa-Krämer}}]{doi:10.1063/1.3060808}%
  \BibitemOpen
  \bibfield  {author} {\bibinfo {author} {\bibfnamefont {M.~A.}\ \bibnamefont
  {Garcia}}, \bibinfo {author} {\bibfnamefont {E.}~\bibnamefont
  {Fernandez~Pinel}}, \bibinfo {author} {\bibfnamefont {J.}~\bibnamefont {de~la
  Venta}}, \bibinfo {author} {\bibfnamefont {A.}~\bibnamefont {Quesada}},
  \bibinfo {author} {\bibfnamefont {V.}~\bibnamefont {Bouzas}}, \bibinfo
  {author} {\bibfnamefont {J.~F.}\ \bibnamefont {Fernández}}, \bibinfo
  {author} {\bibfnamefont {J.~J.}\ \bibnamefont {Romero}}, \bibinfo {author}
  {\bibfnamefont {M.~S.}\ \bibnamefont {Martín~González}}, \ and\ \bibinfo
  {author} {\bibfnamefont {J.~L.}\ \bibnamefont {Costa-Krämer}},\ }\href
  {\doibase 10.1063/1.3060808} {\bibfield  {journal} {\bibinfo  {journal} {J.
  Appl. Phys.}\ }\textbf {\bibinfo {volume} {105}},\ \bibinfo {pages} {013925}
  (\bibinfo {year} {2009})}\BibitemShut {NoStop}%
\bibitem [{\citenamefont {Robinson}\ \emph {et~al.}(2007)\citenamefont
  {Robinson}, \citenamefont {Piano}, \citenamefont {Burnell}, \citenamefont
  {Bell},\ and\ \citenamefont {Blamire}}]{PhysRevB.76.094522}%
  \BibitemOpen
  \bibfield  {author} {\bibinfo {author} {\bibfnamefont {J.~W.~A.}\
  \bibnamefont {Robinson}}, \bibinfo {author} {\bibfnamefont {S.}~\bibnamefont
  {Piano}}, \bibinfo {author} {\bibfnamefont {G.}~\bibnamefont {Burnell}},
  \bibinfo {author} {\bibfnamefont {C.}~\bibnamefont {Bell}}, \ and\ \bibinfo
  {author} {\bibfnamefont {M.~G.}\ \bibnamefont {Blamire}},\ }\href {\doibase
  10.1103/PhysRevB.76.094522} {\bibfield  {journal} {\bibinfo  {journal} {Phys.
  Rev. B}\ }\textbf {\bibinfo {volume} {76}},\ \bibinfo {pages} {094522}
  (\bibinfo {year} {2007})}\BibitemShut {NoStop}%
\bibitem [{\citenamefont {{Kon\u{c}}}\ \emph {et~al.}(1994)\citenamefont
  {{Kon\u{c}}}, \citenamefont {{Spi\u{s}\'{a}k}}, \citenamefont {{Koll\'{a}r}},
  \citenamefont {{Sov\'{a}k}}, \citenamefont {{Du\u{s}a}},\ and\ \citenamefont
  {{Reininger}}}]{312324}%
  \BibitemOpen
  \bibfield  {author} {\bibinfo {author} {\bibfnamefont {M.}~\bibnamefont
  {{Kon\u{c}}}}, \bibinfo {author} {\bibfnamefont {P.}~\bibnamefont
  {{Spi\u{s}\'{a}k}}}, \bibinfo {author} {\bibfnamefont {P.}~\bibnamefont
  {{Koll\'{a}r}}}, \bibinfo {author} {\bibfnamefont {P.}~\bibnamefont
  {{Sov\'{a}k}}}, \bibinfo {author} {\bibfnamefont {O.}~\bibnamefont
  {{Du\u{s}a}}}, \ and\ \bibinfo {author} {\bibfnamefont {T.}~\bibnamefont
  {{Reininger}}},\ }\href {\doibase 10.1109/20.312324} {\bibfield  {journal}
  {\bibinfo  {journal} {IEEE Trans. Magn.}\ }\textbf {\bibinfo {volume} {30}},\
  \bibinfo {pages} {524} (\bibinfo {year} {1994})}\BibitemShut {NoStop}%
\bibitem [{\citenamefont {Bergeret}\ \emph {et~al.}(2001)\citenamefont
  {Bergeret}, \citenamefont {Volkov},\ and\ \citenamefont
  {Efetov}}]{PhysRevB.64.134506}%
  \BibitemOpen
  \bibfield  {author} {\bibinfo {author} {\bibfnamefont {F.~S.}\ \bibnamefont
  {Bergeret}}, \bibinfo {author} {\bibfnamefont {A.~F.}\ \bibnamefont
  {Volkov}}, \ and\ \bibinfo {author} {\bibfnamefont {K.~B.}\ \bibnamefont
  {Efetov}},\ }\href {\doibase 10.1103/PhysRevB.64.134506} {\bibfield
  {journal} {\bibinfo  {journal} {Phys. Rev. B}\ }\textbf {\bibinfo {volume}
  {64}},\ \bibinfo {pages} {134506} (\bibinfo {year} {2001})}\BibitemShut
  {NoStop}%
\bibitem [{\citenamefont {Bergeret}\ and\ \citenamefont
  {Tokatly}(2014)}]{PhysRevB.89.134517}%
  \BibitemOpen
  \bibfield  {author} {\bibinfo {author} {\bibfnamefont {F.~S.}\ \bibnamefont
  {Bergeret}}\ and\ \bibinfo {author} {\bibfnamefont {I.~V.}\ \bibnamefont
  {Tokatly}},\ }\href {\doibase 10.1103/PhysRevB.89.134517} {\bibfield
  {journal} {\bibinfo  {journal} {Phys. Rev. B}\ }\textbf {\bibinfo {volume}
  {89}},\ \bibinfo {pages} {134517} (\bibinfo {year} {2014})}\BibitemShut
  {NoStop}%
\bibitem [{\citenamefont {Glick}\ \emph {et~al.}(2017)\citenamefont {Glick},
  \citenamefont {Edwards}, \citenamefont {Korucu}, \citenamefont {Aguilar},
  \citenamefont {Niedzielski}, \citenamefont {Loloee}, \citenamefont {Pratt},
  \citenamefont {Birge}, \citenamefont {Kotula},\ and\ \citenamefont
  {Missert}}]{PhysRevB.96.224515}%
  \BibitemOpen
  \bibfield  {author} {\bibinfo {author} {\bibfnamefont {J.~A.}\ \bibnamefont
  {Glick}}, \bibinfo {author} {\bibfnamefont {S.}~\bibnamefont {Edwards}},
  \bibinfo {author} {\bibfnamefont {D.}~\bibnamefont {Korucu}}, \bibinfo
  {author} {\bibfnamefont {V.}~\bibnamefont {Aguilar}}, \bibinfo {author}
  {\bibfnamefont {B.~M.}\ \bibnamefont {Niedzielski}}, \bibinfo {author}
  {\bibfnamefont {R.}~\bibnamefont {Loloee}}, \bibinfo {author} {\bibfnamefont
  {W.~P.}\ \bibnamefont {Pratt}}, \bibinfo {author} {\bibfnamefont {N.~O.}\
  \bibnamefont {Birge}}, \bibinfo {author} {\bibfnamefont {P.~G.}\ \bibnamefont
  {Kotula}}, \ and\ \bibinfo {author} {\bibfnamefont {N.}~\bibnamefont
  {Missert}},\ }\href {\doibase 10.1103/PhysRevB.96.224515} {\bibfield
  {journal} {\bibinfo  {journal} {Phys. Rev. B}\ }\textbf {\bibinfo {volume}
  {96}},\ \bibinfo {pages} {224515} (\bibinfo {year} {2017})}\BibitemShut
  {NoStop}%
\end{thebibliography}%

\section*{Acknowledgements}

We wish to thank Norman Birge for advice and helpful discussions, M. Vaughan, J. Massey, M. Rogers, T. Moorsom, M. Ali,  M. Rosamond, and L. Chen for experimental assistance. We acknowledge support from the Henry Royce Institute. The work was supported financially through the following EPSRC grants: EP/M000923/1, EP/P022464/1 and EP/R00661X/1. This project has received funding from the European Unions Horizon 2020 research and innovation programme under the Marie Sk\l{}odowska-Curie Grant Agreement No. 743791 (SUPERSPIN).

\section*{Author Contributions Statement}

N.S. and G.B. conceived and designed the experiment. N.S., P.M.S., E.D., B.J.H. and G.B. undertook the materials development including optimisation and characterisation of the growth process. N.S., T.M. and G.B. undertook the low temperature transport and magnetometry measurements and analysed the data. N.S. undertook the sample fabrication and wrote the manuscript. All authors reviewed and edited the manuscript.

\section*{Additional Information}

\textbf{Supplementary information} accompanies this paper at URL.\\

\noindent \textbf{Competing interests:} The authors declare no competing interests.\\



\begin{table}[ht]
\centering

\begin{tabular}{c|c|c|c|c|c}

\textit{F}          & $\xi_{F1}$   & $\xi_{F2}$     & $d_{\text{zero--}\pi}$   & $V_0$                    & Reference \\ 
                    & (nm)         & (nm)           & (nm)         & ($\mu$V)               &           \\ \hline
Pd$_{97}$Fe$_3$      & $16.2\pm 1.4$          & $7.2\pm 0.6$           & $16.3 \pm 0.2$         &   $102 \pm 12$                 & \cite{7747519}      \\ \hline
Ni$_{80}$Fe$_{20}$ & $1.50 \pm 0.38$ & $0.58 \pm 0.10$   & $1.76 \pm 0.05$ & $69 \pm 19$   & \cite{doi:10.1063/1.4989392}\\ \hline
Ni$_{65}$Fe$_{15}$Co$_{20}$ & $1.11 \pm 0.16$ & $0.48 \pm 0.03$   & $1.15 \pm 0.02$ & $30 \pm 6$     & \cite{doi:10.1063/1.4989392} \\ \hline
Ni$_{73}$Fe$_{21}$Mo$_6$  & $0.48 \pm 0.04$ & $0.955 \pm 0.004$ & $2.25 \pm 0.10$ & $150 \pm 50$   & \cite{Niedzielski_2015}        \\ \hline
Co$_{68}$B$_{32}$  &   $0.28 \pm 0.02$           &   $0.20 \pm 0.02$             &  $0.30\pm0.05$ & $56 \pm 8$              &     This work      \\ 
\end{tabular}
\caption{Best fit parameters determined for selected \textit{S--F--S} Josephson junctions where the \textit{F} layers are ferromagnetic alloys and the $I_cR_N$ oscillations are well described by Equation \ref{eq:intermediate}.\\}
        \label{table1}
\end{table}

\end{document}


\begin{center}
\large{Supplementary information for: ``Pt and CoB trilayer Josephson $\pi$ junctions with perpendicular magnetic anisotropy''}\\
\small{N. Satchell, T. Mitchell, P. M. Shepley, E. Darwin, B. J. Hickey, and G. Burnell\\ School of Physics and Astronomy, University of Leeds, Leeds, LS2 9JT, United Kingdom}
\end{center}



This supplementary information contains additional analysis and data in support of the conclusions presented in the main text. 

\section{Determining Magnetisation}

The magnetisation data presented in the main text were acquired in a Quantum Design MPMS3 magnetometer. The magnetometer returns measurements of the total magnetic moment of the sample in emu. The total magnetic moment of the sample contains the contributions due to the substrate, thin film, and any spurious signal such as those described by Garcia \textit{et al.} \cite{doi:10.1063/1.3060808} . Careful sample handing was used to minimise spurious contributions to the total moment. The linear diamagnetic background due to the substrate was subtracted from the total measured moment. We report the data after substrate subtraction as the area normalised magnetic moment (moment/area) in emu/cm$^2$ by measuring the area of the sample. 

\begin{figure}[] 
\includegraphics[width=0.5\columnwidth]{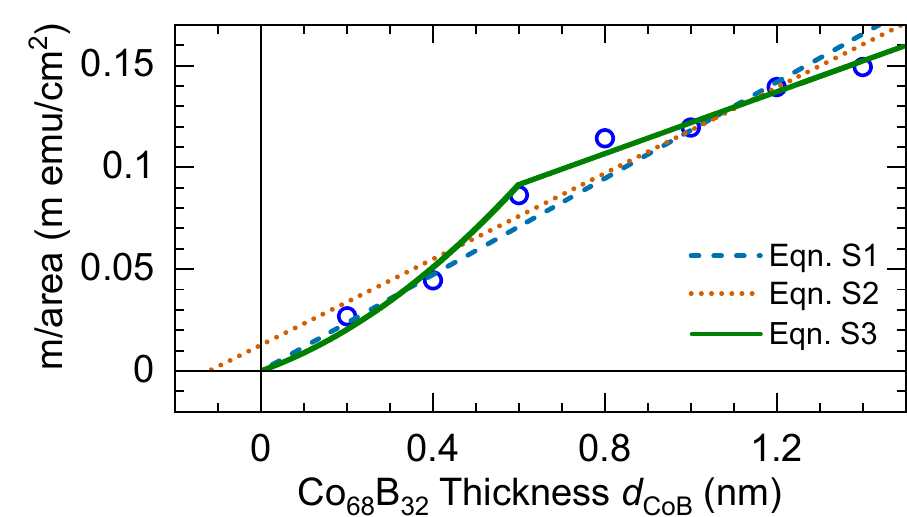} 
\caption{Magnetic moment/area versus Co$_{68}$B$_{32}$ thickness data presented in the main text, fit to Equations \ref{M1}, \ref{M2}, and \ref{M3}. The extracted magnetisation of the Co$_{68}$B$_{32}$ layer from the three fits are $M=1200\pm50$ emu/cm$^3$ (Equation \ref{M1}), $M=1100\pm100$ emu/cm$^3$ (Equation \ref{M2}), and $M=760\pm90$ emu/cm$^3$ (Equation \ref{M3}). Only fitting to Equation \ref{M3} returns a magnetisation consistent with the bulk value of 730 emu/cm$^3$\cite{312324}.}
\label{SM_M1}
\end{figure}

For a uniform slab of ferromagnetic material, the volume magnetisation ($M$) can be determined by dividing the saturation moment/area by the thickness of the ferromagnetic layer. $M$ can also be reliably determined by measuring several samples of varying thickness. If one plots moment/area versus thickness ($d$), as per Figure \ref{SM_M1}, then the fitted gradient will be $M$,
\begin{equation}
\label{M1}
    \text{moment/area}=M  d.
\end{equation} 

In physical systems, a thin ferromagnetic layer is unlikely to be accurately described as a uniform slab. In thin layers the interfaces become significant, so one must account for interfacial contributions to the magnetisation, which may be different than the bulk. Examples of interfacial contributions include magnetic dead layers and the polarization of adjacent layers. Magnetic dead layers can form at interfaces with certain non-ferromagnetic materials or if the surface of the layer becomes oxidised. Of particular relevance to studies of superconductor/ferromagnetc heterostructures is the known magnetic dead layer formation at Nb/ferromagnet interfaces \cite{PhysRevB.76.094522}. Another interfacial effect in thin films is the polarization of the adjacent layer. At some ferromagnet/non-ferromagnet interfaces, the ferromagnetic layer can create a polarization inside the non-ferromagnetic layer by the magnetic proximity effect. Polarization is particularly common at interfaces with Pd and Pt since they are stoner enhanced paramagnets.

To model a magnetic slab with possible magnetic dead layers and/or polarized adjacent layers, one can again plot moment/area versus $d$ and fit to the expression,
\begin{equation}
\label{M2}
    \text{moment/area}= M  (d - d_i),
\end{equation}
\noindent where $d_i$ is the x-axis intercept. A positive x-axis intercept is indicative of magnetic dead layer formation and the thickness of the dead layer at each interface can be estimated as $d_i/2$. A negative x-axis intercept is indicative that at least one of the adjacent layers has gained a polarization. In this instance, the y-intercept provides the magnitude of the contribution of the polarized layer(s) to the overall measured moment/area.  

To describe our Pt/Co$_{68}$B$_{32}$/Pt data, we first attempt fitting to Equations \ref{M1} and \ref{M2}. The results of these fittings are shown in Figure \ref{SM_M1}. We find that these fits overestimate the magnetization of our Co$_{68}$B$_{32}$ layer and underestimate the polarization in the Pt layer. The extracted $M=1200\pm50$ emu/cm$^3$ (Equation \ref{M1}) and $M=1100\pm100$ emu/cm$^3$ (Equation \ref{M2}) are not consistent with the reported literature value of 730 emu/cm$^3$ for Co$_{68}$B$_{32}$ \cite{312324}. It is very well established from the literature that the Pt gains a considerable polarization in thin film multilayers such as ours, which is not accounted for in the fits to Equations \ref{M1} and \ref{M2} presented in Figure \ref{SM_M1}.

A simple slab model is, therefore, insufficient to fully describe our Pt/Co$_{68}$B$_{32}$/Pt system. To fully describe our data on the Pt/Co$_{68}$B$_{32}$/Pt system presented in the main text we create a toy model based on partial layer coverage. For very thin layers (1 or 2 monolayers), it is possible that the layer coverage is not uniform or that the ferromagnet only partially covers the surface. Such a picture is consistent with common thin-film growth modes, where adatoms initially form islands, which coalesce into complete layer coverage as the film thickness increases. Such incomplete layer coverage will affect the polarization of adjacent layers, as the adjacent layer will only gain polarization in the vicinity of the magnetic islands.

To model the partial layer coverage, we assume that: at zero thickness, the layer coverage is 0$\%$; at a critical thickness, $d_\text{critical}$, the islands connect and layer coverage is 100$\%$; layer coverage $\%$ increases linearly between those two thicknesses; and at thicknesses above $d_\text{critical}$, the data should be described by Equation \ref{M2}. The equation describing this toy model is,
\begin{equation}
\label{M3}
\mathrm{moment/area} = \begin{cases} M(d - d_i)(d/d_\text{critical}),&\text{for }0<d<d_\text{critical}\\ 
M(d - d_i),&\text{for }d\ge d_\text{critical} \end{cases}
\end{equation}

\noindent The results fitting to this toy model returns a magnetisation $M=760\pm90$ emu/cm$^3$, which is consistent with the bulk value of 730 emu/cm$^3$\cite{312324}. Equation \ref{M3} therefore best represents our Pt/Co$_{68}$B$_{32}$/Pt system as shown in Figure \ref{SM_M1}, and so this best fit model is presented in the main text.


\section{trapped flux in superconducting coils}

\begin{figure} []
\includegraphics[width=0.5\columnwidth]{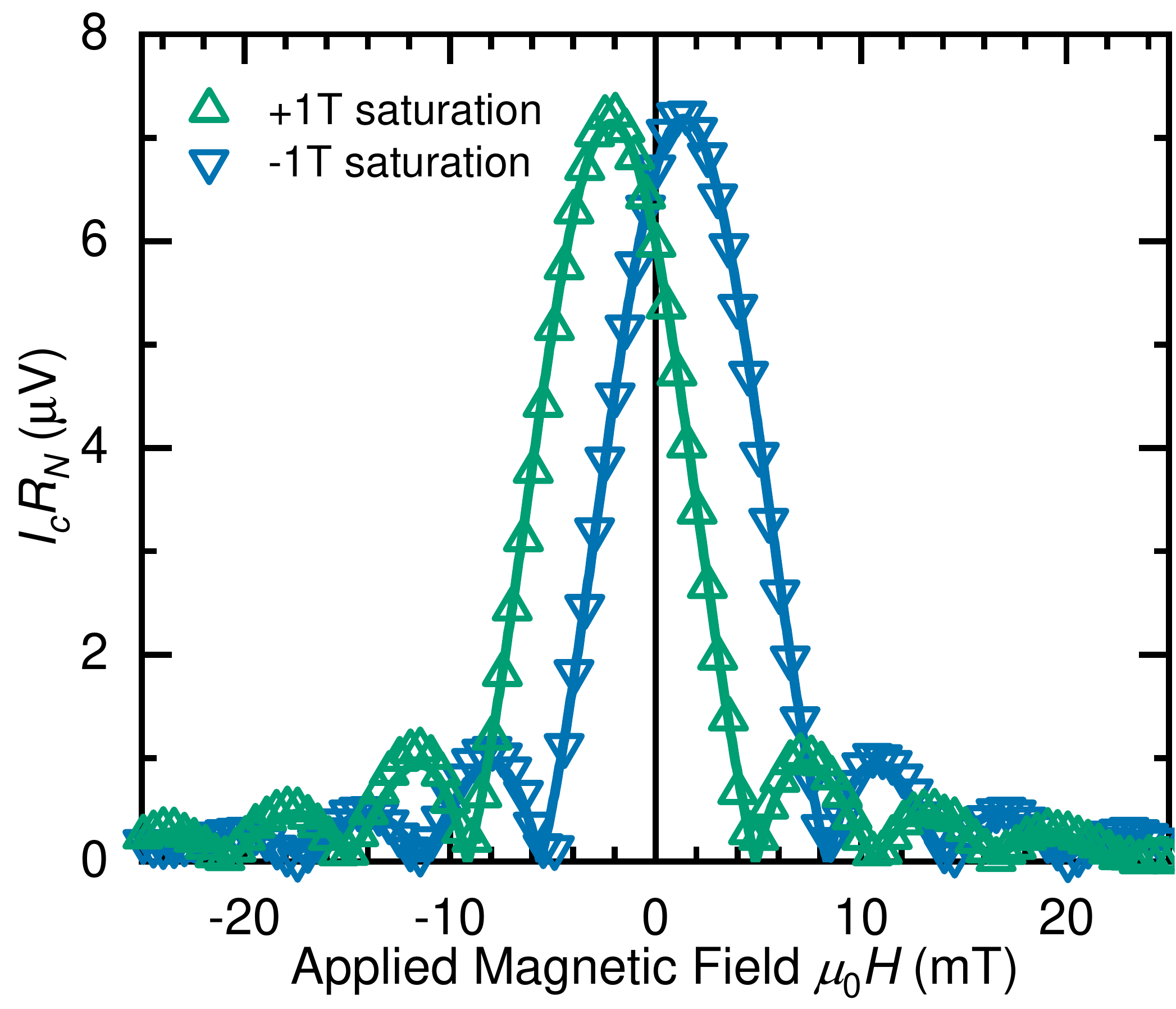} 
\caption{Product of critical Josephson current times normal-state resistance versus applied magnetic field for ferromagnetic Josephson junctions of the form \textit{S}-Pt(10)-Co$_{68}$B$_{32}$(0.6)-Pt(5)-\textit{S} at 1.8~K. $I_c$ is determined from the measured $I$--$V$ characteristic at each field value and $R_N$ is the average normal state resistance across all measured fields. The uncertainty in determining $I_cR_N$ is smaller than the data points. The data are fit to Equation 2 and 3 from the main text.  }
\label{SM3}
\end{figure}

The cryostat has a single horizontal field coil. The sample can be rotated about the vertical axis, which we perform in increments of 90$^{\circ}$ to bring that field in- and out-of-plane of the junctions. In order to avoid trapping flux in the superconducting Nb layers in the devices, we performed all sample rotations in zero field above the $T_c$ of the Nb and always cooled the sample in zero applied field (in practice there will inevitably be a small remanent field due to trapped flux in the magnet). The full sequence of setting the magnetic state of our samples and performing the measurement field sweeps is: warm to 15K, rotate sample to apply field out-of-plane, apply saturating field, remove saturating field, rotate sample by 90$^{\circ}$, cool sample, apply field in-plane, measure. Once we have finished measuring that magnetic state of the sample, we may wish to measure a further condition, such as reversing the magnetisation. To do so, we remove the in-plane field, warm to 15K and repeat the cycle described above.

Figure \ref{SM3} contains supporting data on the trapped flux in the superconducting coils used in the measurements of our samples. We have observed similar trends by measuring the field directly using a Hall probe during magnetic field sweeps when the sample space was at room temperature. The green data are acquired by sweeping from +25~mT to -25~mT after performing the cycle described above with a +1~T out-of-plane saturating field at 15~K. The blue data are acquired by sweeping from -25~mT to +25~mT after cycling through a -1~T out-of-plane saturating field at 15~K. These data suggest that to compensate for the trapped flux in our 3~T horizontal superconducting Helmholtz coils to achieve zero global applied field requires application of a small opposite field of 1-3~mT.

\section{Goodness of fit analysis}

\begin{figure}[] 
\includegraphics[width=0.9\columnwidth]{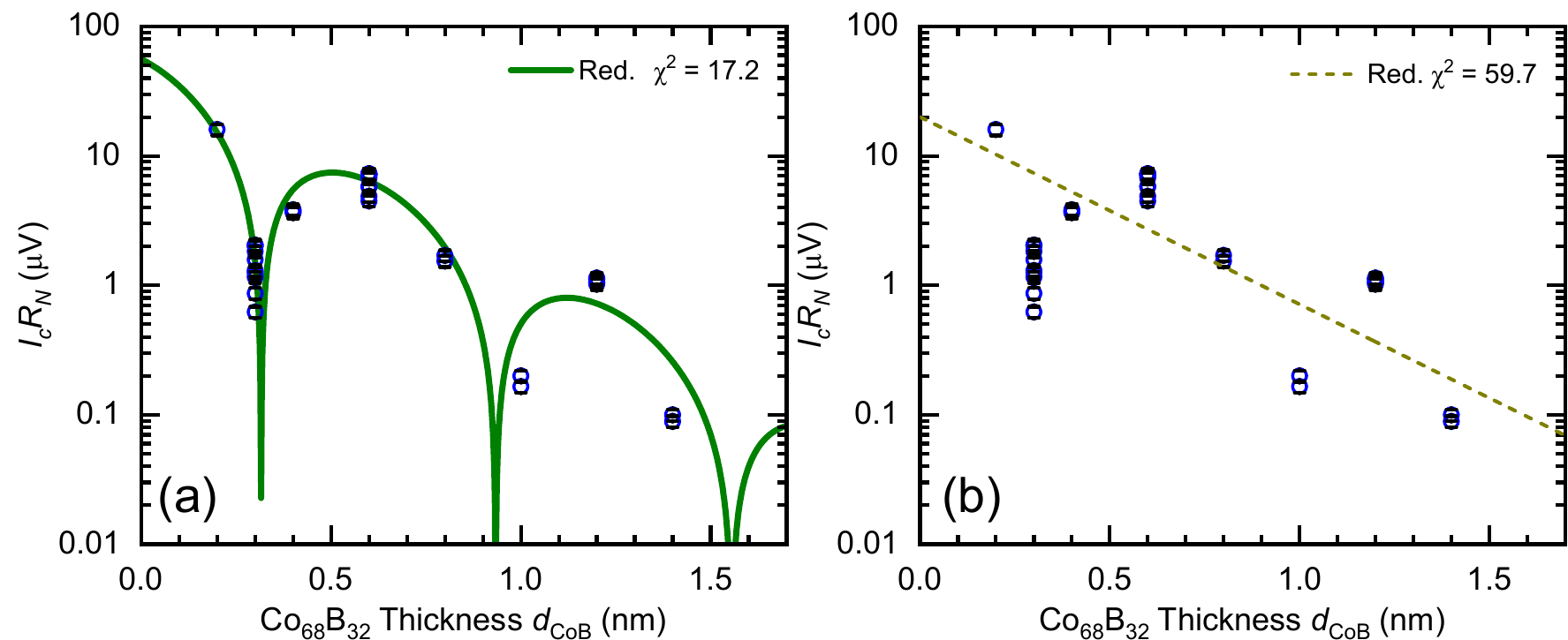} 
\caption{Product of critical Josephson current times normal-state resistance versus nominal Co$_{68}$B$_{32}$ thickness for ferromagnetic Josephson junctions of the form \textit{S}-Pt(10)-Co$_{68}$B$_{32}$(\textit{d}$_\text{CoB}$)-Pt(5)-\textit{S}. Each data point represents one Josephson junction. (a) The best fit of Equation \ref{eq:intermediate} as presented in the main text of our publication, where the reduced $\chi^2 = 17.2$. (b) The best fit of Equation \ref{null}, the null hypothesis, where $\xi_{F} = 0.30$~nm and the reduced $\chi^2 = 59.7$.}
\label{sm1}
\end{figure}

\begin{figure} [h]
\includegraphics[width=0.9\columnwidth]{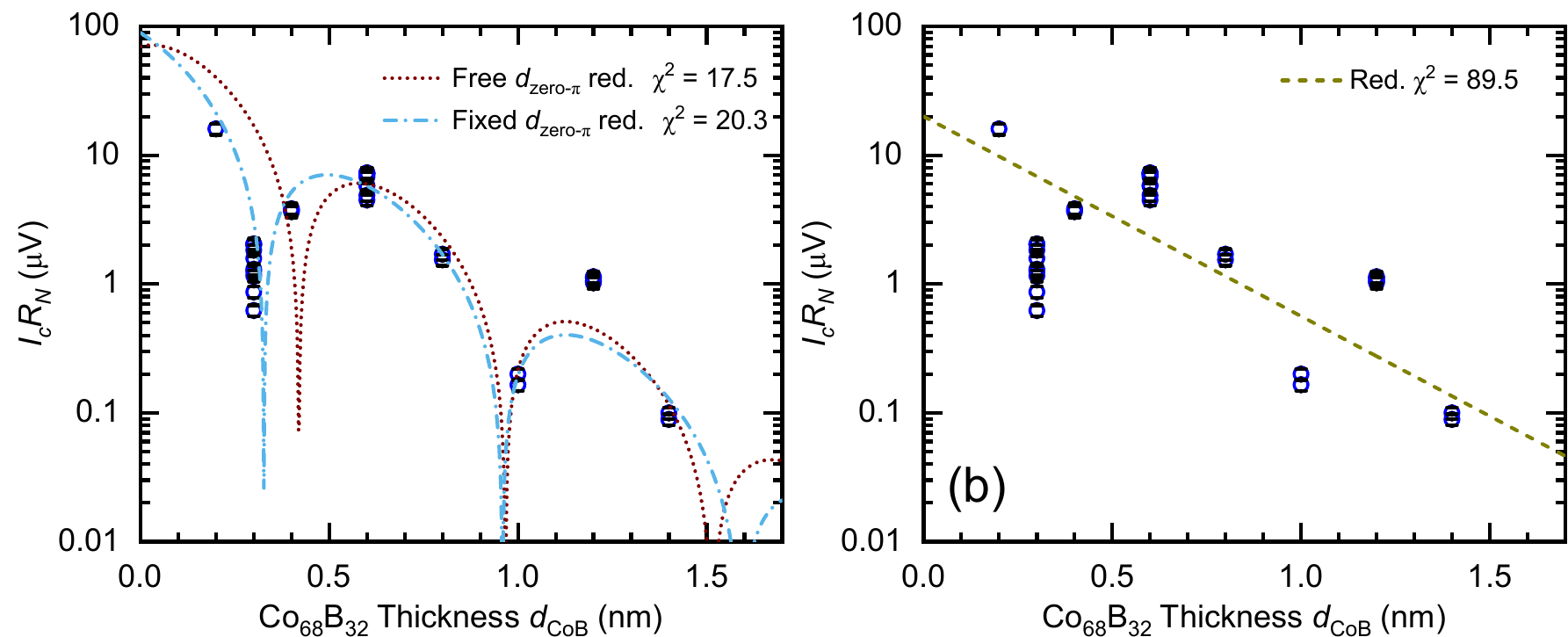} 
\caption{Product of critical Josephson current times normal-state resistance versus nominal Co$_{68}$B$_{32}$ thickness for ferromagnetic Josephson junctions of the form \textit{S}-Pt(10)-Co$_{68}$B$_{32}$(\textit{d}$_\text{CoB}$)-Pt(5)-\textit{S}. Each data point represents one Josephson junction. In this case, fitting is only performed for the data range $d_\text{CoB} \geq 0.6$~nm. (a) Fitting to Equation \ref{eq:intermediate} with either a free or fixed $d_{\text{zero}-\pi}$ parameter. The trends in the fitted parameters over this reduced data range are consistent with fitting over the entire data range. (b) Fit to Equation \ref{null}, where $\xi_{F} = 0.28$~nm and the reduced $\chi^2 = 89.5$.}
\label{SM2}
\end{figure}

The main conclusion of our work is that the transport properties of Josephson junctions with Pt/Co$_{68}$B$_{32}$/Pt barriers are consistent with intermediate limit zero-$\pi$ oscillations. For convenience, we reproduce the equation governing this limit \cite{PhysRevB.64.134506},
\begin{equation}\label{eq:intermediate}
I_cR_N = V_0 \exp \bigg( \frac{-d_F}{\xi_{F1}} \bigg) \bigg|\sin\bigg( \frac{d_F - d_{\text{zero--}\pi}}{\xi_{F2}} \bigg)\bigg|,
\end{equation}
\noindent where $V_0$ is the extrapolated $I_cR_N$ at zero thickness, $d_{\text{zero--}\pi}$ is the position of the first zero--$\pi$ transition, $\xi_{F1} = l_e$ and $\xi_{F2}=\xi_{F}$ are the lengthscales governing the decay and oscillation of $I_cR_N$, respectively. For ferromagnetic Josephson junctions where zero-$\pi$ oscillations are not expected to occur, the decay of supercurrent can be described by the simple exponential decay,
\begin{equation}
\label{null}
I_cR_N = V_0 \exp \big(-d_F/\xi_F \big).
\end{equation}

\noindent A possible null hypothesis for our reported zero-$\pi$ oscillations in Pt/Co$_{68}$B$_{32}$/Pt barriers is if our junctions are best described by Equation \ref{null} and not Equation \ref{eq:intermediate}, as this would suggest that zero-$\pi$ oscillations may not have been observed. 

Figure \ref{sm1} shows goodness of fit analysis of our result compared to the null hypothesis. For Figure \ref{sm1} (a), the best fit of Equation \ref{eq:intermediate} as presented in the main text gives the reduced $\chi^2 = 17.2$. For Figure \ref{sm1} (b), the best fit of Equation \ref{null} gives the reduced $\chi^2 = 59.7$. Therefore, the $\chi^2$ analysis allows us to reject the null hypothesis that a simple exponential decay can describe our experimental result.

In order to fully describe our magnetisation data, we constructed a toy model of partial layer coverage for samples with the thinnest layers. On the other hand, electrical transport measurements on junctions with the thinnest barriers do not appear to be sensitive to localised partial coverage, as the pair correlations which probe the barrier are averaged over the superconducting coherence length. Nonetheless, we consider next if our findings are robust even when excluding the data from the thinnest barriers. 

Figure \ref{SM2} shows goodness of fit analysis on a reduced data range $d_\text{CoB} \geq 0.6$~nm. In Figure \ref{SM2}  (a) we consider the results of two fits to Equation \ref{eq:intermediate}. For the first fit attempt, where all parameters in Equation \ref{eq:intermediate} are free parameters (dotted line), the fitting struggles to fit the $d_{\text{zero}-\pi}$ parameter, as the reduced data range does not contain the first zero-$\pi$ transition. However overall, the trends in the fitted $\xi_{F1} = 0.22\pm0.02$~nm and $\xi_{F2} = 0.17\pm0.02$~nm are consistent with our conclusion of intermediate limit transport. Alternatively, we can fix the $d_{\text{zero}-\pi}$ parameter (dot dash line), in which case the best fit returns $\xi_{F1} = 0.22\pm0.02$~nm and $\xi_{F2} = 0.20\pm0.01$~nm. By eye, this fit is a plausible explanation of our results over the entire data range. The returned $\xi_{F2}$ values from these two fitting methodologies, where $\xi_{F2}$ governs the period of the zero-$\pi$ oscillation (the key result of our work), are consistent, within error, with  $\xi_{F2}$ by fitting over the entire data range ($0.20\pm0.02$~nm), providing us confidence in our reported result. In Figure \ref{SM2} (b) we fit to the simple exponential decay Equation \ref{null}, over the reduced data range, and find $\xi_{F} = 0.28$~nm and the reduced $\chi^2 = 89.5$. The $\chi^2$ analysis allows us to reject the null hypothesis that a simple exponential decay can describe our experimental result, even over the reduced data range $d_\text{CoB} \geq 0.6$~nm.

\section{Possible Spin-Triplet Component of supercurrent}


Josephson junctions containing a source of $s$-wave superconductivity, large spin-orbit coupling, and ferromagnetism are predicted to display transport properties consistent with spin-triplet supercurrents \cite{PhysRevB.89.134517}. Assuming that the total supercurrent in the junction is the sum of the spin-singlet and spin-triplet components, we can describe the total supercurrent as;
\begin{equation}\label{eq:triplet}
I_cR_N = V_0 \exp \bigg( \frac{-d_F}{\xi_{F1}} \bigg) \bigg|\sin\bigg( \frac{d_F - d_{\text{zero--}\pi}}{\xi_{F2}} \bigg)\bigg| +  V_\text{Triplet} \exp \bigg( \frac{-d_F}{\xi_\text{Triplet}} \bigg),
\end{equation}
\noindent where $V_\text{Triplet}$ and $\xi_\text{Triplet}$ are the critical voltage and decay length of the spin-triplet component of supercurrent, respectively. 

\begin{figure}[] 
\includegraphics[width=0.5\columnwidth]{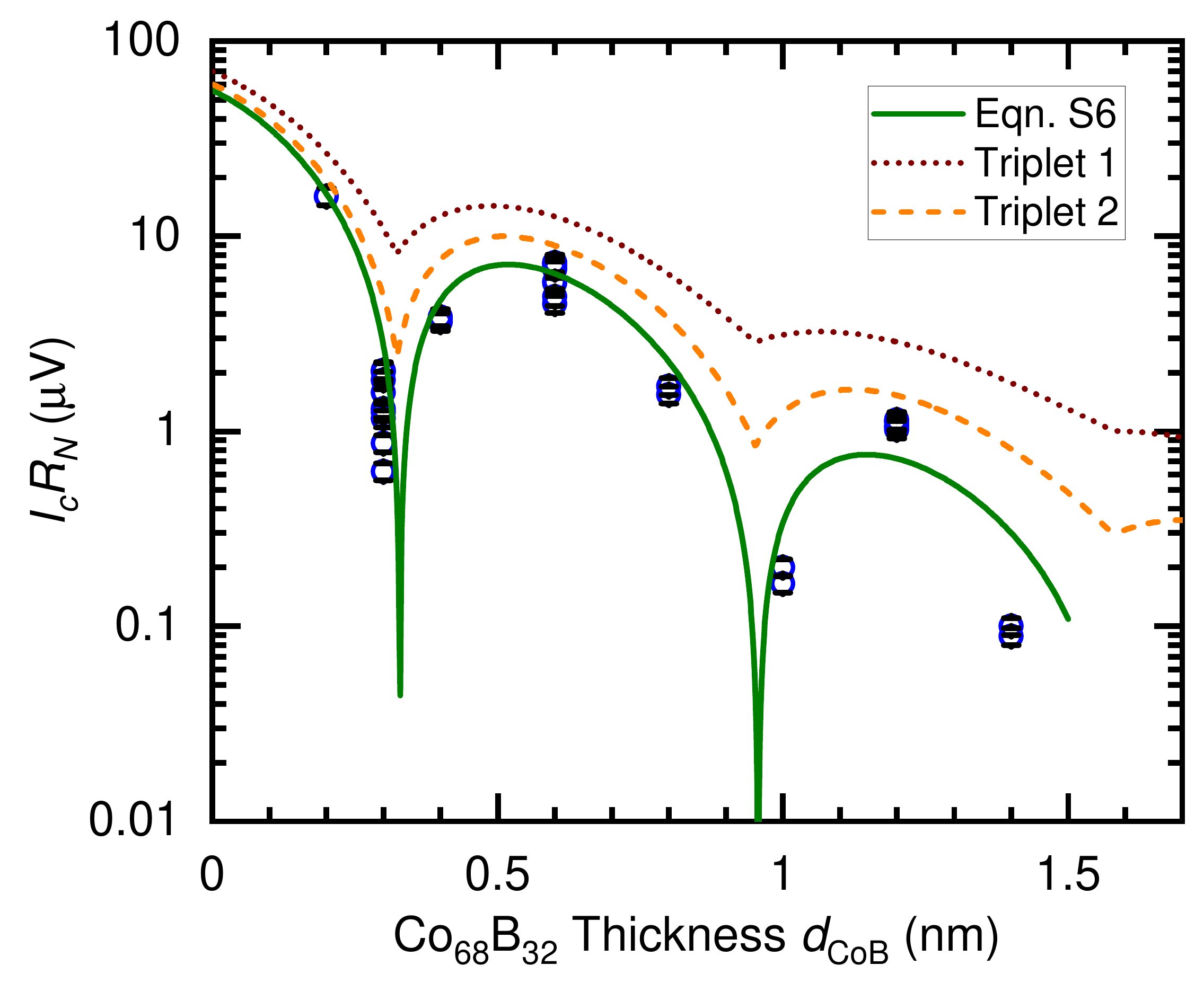} 
\caption{Product of critical Josephson current times normal-state resistance versus nominal Co$_{68}$B$_{32}$ thickness for ferromagnetic Josephson junctions of the form \textit{S}-Pt(10)-Co$_{68}$B$_{32}$(\textit{d}$_\text{CoB}$)-Pt(5)-\textit{S}. Each data point represents one Josephson junction. The solid green curve shows the best fit of Equation \ref{eq:triplet}, which recovers the fit presented in the main text of our publication as $V_\text{Triplet} \to 0$. The dotted and dashed lines correspond to the simulations of Equation \ref{eq:triplet} to two parameter sets described in the text}
\label{TripletFig}
\end{figure}

To determine if spin-triplet supercurrents contribute to our transport results, we fit Equation \ref{eq:triplet} to our junction data, shown in Figure \ref{TripletFig}. We find that $V_\text{Triplet} \to 0$, and the other fit parameters correspond to those of the singlet only fit described in the main text. This suggests that the data are well described by singlet transport physics alone and any spin-triplet component is below our experimental sensitivity. 

An alternative methodology is to consider whether we might be able to observe a spin-triplet supercurrent of the magnitude reported in literature. Figure \ref{TripletFig} shows simulations of Equation \ref{eq:triplet} with a weak spin-triplet component coexisting in our junctions, inspired by the findings of Glick \textit{et al.} for a PMA synthetic antiferromagnet system with intentional spin-mixing layers \cite{PhysRevB.96.224515}. Parameter set `Triplet 1' corresponds to a decay length of spin-triplet component being only twice that of the spin-singlet component and $V_\text{Triplet} = \frac{1}{7} V_0$. These parameters correspond to the results of Glick \textit{et al.} with symmetric Ni spin-mixer layers. Parameter set `Triplet 2' corresponds to a decay length of spin-triplet component being twice that of the spin-singlet component and $V_\text{Triplet} = \frac{1}{21} V_0$. These parameters correspond to asymmetric Ni and NiFe spin-mixer layers. These simulations suggest that a spin-triplet supercurrent of the magnitude reported in PMA synthetic antiferromagnets by Glick \textit{et al.} should be observable as the major suppression of the oscillations and higher $I_cR_N$ in the thicker samples in this work. 





\bibliography{library}